\documentclass[12pt,onecolumn,twoside]{IEEEtran}

\usepackage[utf8]{inputenc}
\usepackage[T1]{fontenc}
\usepackage{url}              
\usepackage{cite}             

\usepackage[cmex10]{amsmath}  
\usepackage{mleftright}       
\mleftright                   

\usepackage{graphicx}         
\usepackage{booktabs}         
\usepackage{bm}

\usepackage{amsmath}
\usepackage{latexsym}
\usepackage{amssymb}

\newcommand{\cA}{{\cal A}}
\newcommand{\cB}{{\cal B}}
\newcommand{\cC}{{\cal C}}

\newcommand{\cF}{{\cal F}}

\newcommand{\cS}{{\cal S}}
\newcommand{\cT}{{\cal T}}

\newcommand{\mmod}{{\mbox{mod}}}

\newcommand{\blds}{{\mathbf{s}}}

\newcommand{\field}[1]{\mathbb{#1}}

\newcommand{\C}{\field{C}}

\newcommand{\F}{\field{F}}

\newcommand{\bF}{\mathbb{F}}
\newcommand{\Tr}{\mathrm{Tr}}

\newtheorem{theorem}{Theorem}
\newtheorem{definition}{Definition}
\newtheorem{lemma}[theorem]{Lemma}
\newtheorem{corollary}[theorem]{Corollary}
\newtheorem{proposition}[theorem]{Proposition}
\newtheorem{example}{Example}

\newtheorem{conjecture}{Conjecture}

\begin{document}

\title{On de Bruijn Array Codes,\\Part II: Linear Codes\vspace{-1.0ex}}

\date{\today}
\author{\textbf{Simon R. Blackburn$^\text{+}$}, \textbf{Yeow Meng Chee$^\text{x}$}, \textbf{Tuvi Etzion$^{*}$}, \textbf{Huimin Lao$^{\$}$}\\
{\small $^\text{+}$Dept. of Mathematics, Royal Holloway, University of London, Egham, Surrey TW20 0EX, United Kingdom}\\
{\small $^\text{x}$Singapore University of Technology and Design, Singapore}\\
{\small $^*$Computer Science Department, Technion, Israel Institute of Technology, Haifa 3200003, Israel}\\
{\small $^{\$}$School of Physical and Mathematical Sciences, Nanyang Technological University, Singapore}\\
{\small {\it s.blackburn@rhul.ac.uk}, \small {\it ymchee@sutd.edu.sg}, {\it etzion@cs.technion.ac.il}, {\it huimin.lao@ntu.edu.sg}\vspace{-0.13ex}}
\thanks{Parts of this work have been presented at the \emph{IEEE International Symposium on Information Theory 2024}, Athens, Greece, July 2024.
Parts of this work have been presented at the \emph{IEEE International Symposium on Information Theory 2025}, Ann Arbor, Michigan, USA, June 2025.
The research of Tuvi Etzion was supported in part by the Israeli Science Foundation grant no. 222/19.}
}

\maketitle

\begin{abstract}
An M-sequence generated by a primitive polynomial has many interesting and desirable properties.
A pseudo-random array is the two-dimensional generalization of an M-sequence.
There are non-primitive polynomials all of whose non-zero sequences have the same period. These polynomials
generate \emph{sets} of sequences with properties similar to M-sequences.
In this paper, a two-dimensional generalization for such sequences is given. This generalization is for a pseudo-random
array code, which is a set of $r_1 \times r_2$ arrays in which each $n_1 \times n_2$ nonzero matrix is
contained exactly once as a window in one of the arrays. Moreover, these arrays have the shift-and-add property, i.e., the bitwise
addition of two arrays (or a nontrivial shift of such arrays) is another array (or a shift of another array) from the code.
All the known arrays can be formed by folding sequences generated from an irreducible polynomial or a reducible
polynomial whose factors have the same degree and the same exponent. Two proof techniques are used to
prove the constructions are indeed of pseudo-random array codes. The first technique is based on another method,
different from folding, for constructing some of these arrays. The second technique is a generalization of a known
proof technique. This generalization enables the construction of
pseudo-random arrays with parameters not known before, and also provides a variety of pseudo-random array
codes which cannot be generated by the first method. The two techniques also suggest two different hierarchies between pseudo-random array codes.
Finally, two methods to verify whether a folding of sequences, generated by these polynomials, yields a pseudo-random array or a pseudo-random array
code, will be presented.
\end{abstract}

\section{Introduction}
\label{sec:PM+PR}

Generalizations of one-dimensional sequences and codes to higher dimensions are quite natural and fashionable from
both theoretical and practical points of view. Such generalizations were considered
for various structures such as error-correcting codes~\cite{BlBr00,Rot91}, burst-correcting codes~\cite{BBV,BBZS,EtVa02,EtYa09},
constrained codes~\cite{ScBr08,TER09}, and de Bruijn sequences~\cite{Etz88,Mit95,Pat94}.
This paper considers a generalization of one-dimensional linear sequences with a window property to two-dimensional linear arrays with a window
property. For simplicity, only binary arrays and binary sequences are considered in this paper, although some of the
results can be generalized to any alphabet of a finite field $\F_q$ with $q$ elements.

A~{\bf \emph{span~$n$ de Bruijn sequence}} is a cyclic sequence of length $2^n$ in which each $n$-tuple is contained in
exactly one window of $n$ consecutive digits. A {\bf \emph{span $n$ shortened de Bruijn sequence}} is a sequence of length $2^n-1$ in which
each nonzero $n$-tuple is contained in exactly one window of $n$ consecutive digits.
A span $n$ M-sequence (pseudo-random sequence) $\cS$ is a span $n$ shortened de Bruijn sequence such that if $\cS'$ is a nontrivial shift
of $\cS$, then $\cS + \cS'$ is another nontrivial shift of $\cS$. This is the {\bf \emph{shift-and-add property}} of an M-sequence.
The following definition generalizes the definition of a de Bruijn sequence to a two-dimensional array.

\begin{definition}
A {\bf \emph{de Bruijn array}} (often known as a {\bf \emph{perfect map}}) is an $r_1 \times r_2$ doubly-periodic array
(cyclic horizontally and vertically like in a torus),
such that each $n_1 \times n_2$ binary matrix appears exactly once as a window in the array.
\end{definition}

Perfect maps were first presented by Reed and Stewart~\cite{ReSt62} and considered in hundreds of papers
which discussed constructions and applications of these arrays.
The definition was generalized in~\cite{Etz24a,Etz24b} for a set of arrays, i.e., a code which contains
arrays of the same dimensions, as follows.

\begin{definition}
A {\bf \emph{de Bruijn array code}} is a set of $r_1 \times r_2$ doubly-periodic arrays,
such that each $n_1 \times n_2$ matrix appears exactly once as a window in one of the arrays.
Such a set of arrays will be referred to as $(r_1,r_2;n_1,n_2)$-DBAC.
\end{definition}

In general, the {\bf {\emph size}} size $\Delta$ of an array code is the number of arrays (codewords) in the code.
In the previous part of this work~\cite{Etz24b}, several direct and recursive constructions for de Bruijn arrays codes are presented.

\begin{definition}
A {\bf \emph{shortened de Bruijn array}} (or a {\bf \emph{shortened perfect map}}) is an $r_1 \times r_2$
doubly-periodic array, such that $r_1 r_2=2^{n_1 n_2}-1$ and each nonzero $n_1 \times n_2$ matrix appears exactly
once as a window in the array. Such an array will be called an $(r_1,r_2;n_1,n_2)$-SDBA.
A {\bf \emph{shortened de Bruijn array code}} $\C$ is a set of $r_1 \times r_2$ arrays such that $r_1 r_2$ divides $2^{n_1n_2}-1$
and each nonzero $n_1 \times n_2$ matrix is contained exactly once in one of the arrays. Such a code
will be denoted by an $(r_1,r_2;n_1,n_2)$-SDBAC.
\end{definition}

The following lemma is an immediate consequence from the definition.
\begin{lemma}
\label{lem:condSPM}
If $\C$ is an $(r_1,r_2;n_1,n_2)$-$\textup{SDBAC}$ of size $\Delta$, then
\begin{enumerate}
\item[{\bf 1.}] $r_1 > n_1$ or $r_1=n_1=1$,

\item[{\bf 2.}] $r_2 > n_2$ or $r_2=n_2=1$, and

\item[{\bf 3.}] $\Delta r_1r_2 = 2^{n_1n_2} -1$.
\end{enumerate}
\end{lemma}

\begin{definition}
An  $(r_1,r_2;n_1,n_2)$ {\bf \emph{pseudo-random array}} (referred to as $(r_1,r_2;n_1,n_2)$-PRA for short) $\cA$ is an $(r_1,r_2;n_1,n_2)$-SDBA,
for which if $\cA'$ is a nontrivial shift (both horizontal and vertical shifts, where at least one of them is not trivial)
of $\cA$, then $\cA + \cA'$ is also a shift of $\cA$,
where the addition is performed bit-by-bit. This is the {\bf \emph{shift-and-add property}} of the PRA.
\end{definition}

Pseudo-random arrays were constructed first in~\cite{NMIF72,McSl76,Spa63}
and they (along with perfect maps) have found various important applications.
They were used in pattern recognition for structured light systems as described in Geng~\cite{Gen11},
Lin et al.~\cite{LNS16}, Morano et al.~\cite{MOCDZN98},
Salvi et al.~\cite{SFPL10}, and Salvi et al.~\cite{SPB04}.
They are also used in transforming a planar surface into a sensitive touch screen display, see Dai and Chung~\cite{DaCh14}, in
camera localization as described by Szentandrasi et al.~\cite{SZHHDK}, and
in one-shot shape acquisition, e.g., Pag\`{e}s et al.~\cite{PSCF05}. Finally, they can be applied
to surface measurements as described in Kiyasu et al.~\cite{KHYF95} and in Spoelder et al.~\cite{SVPG00},
and also in coded aperture imaging as was presented for example in Gottesman and Fenimore~\cite{GoFe89}.

\begin{definition}
An $(r_1,r_2;n_1,n_2)$ {\bf \emph{pseudo-random array code}}  ($(r_1,r_2;n_1,n_2)$-PRAC for short) $\C$ is an $(r_1,r_2;n_1,n_2)$-SDBAC
for which if $\cA, \cB \in \C$ ($\cA$ and $\cB$ are either distinct or $\cB$ is a nontrivial shift of~$\cA$), then $\cA + \cB$ is also a codeword of $\C$.
This is the {\bf \emph{shift-and-add property}} of the pseudo-random array code.
\end{definition}

The shift-and-add property of a pseudo-random array code implies that its codewords and all their possible cyclic shifts horizontally and vertically
together with the $r_1 \times r_2$ all-zero matrix form a linear array code.

In the current work, constructions for pseudo-random arrays and pseudo-random array
codes are presented. The codes are formed by folding nonzero
sequences generated by a polynomial whose nonzero sequences have the same length.
This technique is well known, but it is not known which folded sequences yield such arrays and array codes.
Two proof techniques are used to prove the correctness
of the suggested construction methods. The first one is based on the
linear span of the bitwise product of sequences generated by two polynomials;
the non-zero sequences produced by each polynomial must all have the same length.
The second proof technique is a direct generalization to a proof method suggested by
MacWilliams and Sloane~\cite{McSl76}. This proof technique implies pseudo-random arrays with parameters which were not obtained
in~\cite{McSl76}, and many pseudo-random array codes which cannot be obtained using the first proof technique.
There are many codes which can be obtained using both proof techniques, but each technique gives many codes which are not obtained by the other technique.

The rest of this paper is organized as follows. In Section~\ref{sec:prelim} we present the basic definitions and theory for
the arrays considered here. In particular the section introduces some theory of linear shift-register sequences and their associated
polynomials. Special attention is given to those polynomials whose nonzero sequences have the same length.
These sequences will be the building blocks of our constructions.
Two operators on these sequences are defined and explained; these operators play an important role in the proofs of correctness.
Also an isomorphism between elements in a finite field and related sequences will be explained in this section, which will be used in
the generalization of the proof technique presented in~\cite{McSl76}.
Section~\ref{sec:folding} introduces the main technique in this paper, known as folding.
This technique is used to form a PRA, and a proof for PRAs with new parameters is given in this section.
Section~\ref{sec:foldZ} generalizes the folding technique for many sequences
to form PRACs. Surprisingly, there is a different method to generate some of these codes, which
suggests another way to prove the correctness of the construction.
However, codes with new parameters are also introduced by this method.
Section~\ref{sec:sufficient} presents a necessary and sufficient condition that the folding technique produces a PRA or a PRAC.
The method presented is applied only on sequences obtained from irreducible polynomials.
The technique is a generalization of a method introduced by Lempel and Cohn~\cite{LeCo85} for generating
sequences for VLSI testing. Our generalization also yields more sequences that can be used in VLSI testing.
The necessary and sufficient condition will be applied to examine if the arrays and the codes
that were constructed by folding sequences generated by irreducible polynomials are PRAs and PRACs, respectively.
The technique can verify whether such array is a pseudo-random array, or whether a set of arrays is a pseudo-random array code.
However, the technique requires more computations to verify the structure than the other two techniques.
Section~\ref{sec:irreducible} introduces the proof technique for the constructed array codes.
Section~\ref{sec:foldIrr} presents a generalization of the proof technique used in Section~\ref{sec:folding}
to prove the existence of more PRACs based on folding.
Folding can be used with sequences of various types of polynomials (primitive, irreducible and not primitive, and reducible)
for which the generated nonzero sequences have the same length.
Analysis of the parameters of PRACs obtained by the various constructions is given
in Section~\ref{sec:analysis}. This analysis suggests a hierarchy between the various codes based on the various
types of polynomials used in their construction.
A second hierarchy for $r_1 \times r_2$ arrays codes obtained by folding is by containment of their window size.
A comparison between the array codes obtained by the two proof techniques is given in this section.
Section~\ref{sec:second_Necc_Suff} contains another necessary and sufficient condition to verify whether the array or code obtained
from folding is a PRA or a PRAC, respectively.
This condition is more computationally efficient (for larger parameters) than the technique given in Section~\ref{sec:sufficient}.
A conclusion and discussion of further research is given in Section~\ref{sec:conclude}.

\section{Preliminaries}
\label{sec:prelim}

We start this section with the theory of shift-register sequences.
Shift registers and their sequences have been extensively studied and their theory can be
found in the following comprehensive books:~\cite{Etz24,Gol67,GoKl12,LiNi97}.
The material in this section is presented in these books. This theory will be applied later to our two-dimensional arrays.

\begin{figure}[ht]
\vspace{0.4cm}
\begin{picture}(115,115)(-70,50)
\includegraphics[width=10cm]{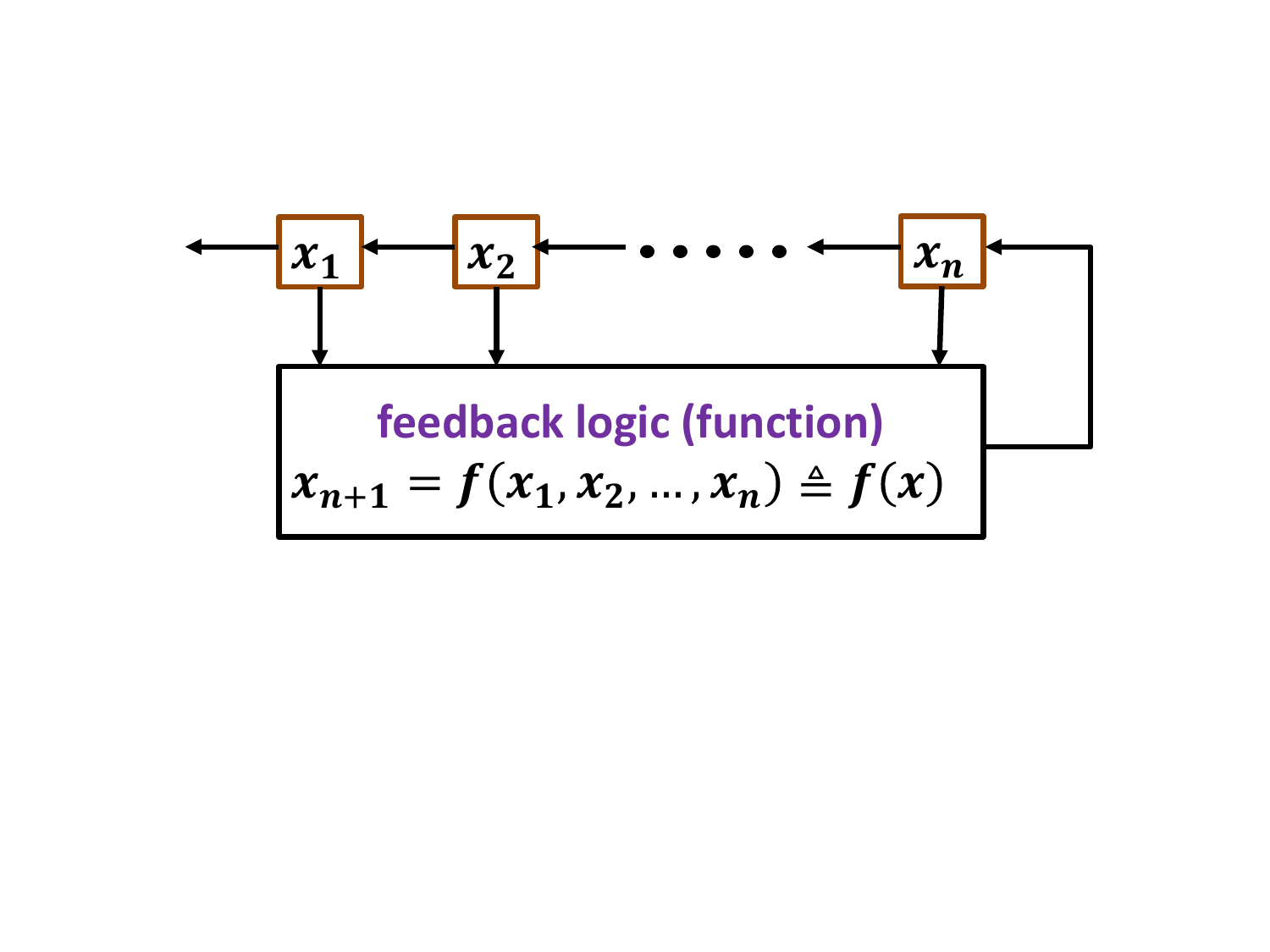}
\end{picture}
\vspace{-0.3cm}
\caption{Feedback shift register of order $n$.}
\label{fig:FSRn}
\end{figure}

A {\bf \emph{feedback shift register}} of order $n$ (an FSR$_n$ in short) has $2^n$ {\bf \emph{states}}, represented by the set
of $2^n$~binary words of length $n$.
An FSR$_n$ has $n$ cells, which are binary storage elements,
where each cell stores at each stage one of the bits of the current state $x=(x_1,x_2,\ldots,x_n)$.
Such an FSR$_n$ is depicted in Fig.~\ref{fig:FSRn}.

If the word $(x_1,x_2, \ldots , x_n)$ is the state of the FSR$_n$, then $x_i$ is stored in the $i$-th cell of the FSR$_n$.
The $n$~cells are connected to another logic element which computes a Boolean function
$f(x_1,x_2,\ldots,x_n)$. At periodic intervals, controlled and synchronized by a global clock, $x_2$ is transferred to $x_1$,
$x_3$ to $x_2$, and so on until $x_n$ is transferred to $x_{n-1}$. The value of the feedback function is
transferred to $x_n$. Hence, it is common to write $x_{n+1}=f(x_1,x_2,\ldots,x_n)$.
The register starts with an {\bf \emph{initial state}} $(a_1,a_2,\ldots,a_n)$, where $a_i$, $1 \leq i \leq n$,
is the initial value stored in the $i$-th cell.

%

A {\bf \emph{linear feedback shift register}} of order $n$ (LFSR$_n$ in short) is a shift register whose feedback function is linear.
We are interested only in nonsingular LFSRs. Such an LFSR$_n$ has a feedback function of the form
$$
x_{n+1}=f(x)=f(x_1,x_2,\ldots,x_n)=\sum_{i=1}^n c_i x_{n+1-i}, ~~~ c_i \in \{0,1\}, ~ 1 \leq i \leq n-1, ~ c_n=1.
$$

The associated LFSR$_n$ sequence $(a_k)_{k=-n}^\infty$, where the initial state is $(a_{-n},a_{-n+1},\ldots,a_{-1})$, satisfies a linear recursion
\begin{equation}
\label{eq:recur_seq}
a_k = \sum_{i=1}^n c_i a_{k-i} , ~~~ k=0,1,\ldots .
\end{equation}

The {\bf \emph{characteristic polynomial}} of the sequence $( a_k)$ is defined by
\begin{equation*}
\label{eq:char_poly}
c(x)=1-\sum_{i=1}^n c_i x^i .
\end{equation*}
We say that the (characteristic) polynomial $c(x)$ {\bf \emph{generates}} the (infinite) sequence $( a_k)$.
Any such polynomial can be written in a compact way as $c_n c_{n-1} ~ \cdots c_1 1$. This sequence might be cyclic and be
also represented differently.

The sequences generated by various characteristic polynomials are used here to form the PRACs. Hence,
we will define a few properties of these sequences and a few operations on them. These properties and operations
will be used to construct PRACs or to prove the correctness of the constructions.

We distinguish between acyclic and cyclic sequences.
A {\bf \emph{cyclic sequence}} denoted by $[a_0 a_1 \cdots a_{n-1}]$ and an {\bf \emph{acyclic sequence}} (a word) denoted by $(a_0 a_1 \cdots a_{n-1})$.
An acylic sequence has a starting position. In contrast, we think of different shifts of a cyclic sequence as being equal.
Cyclic sequences can sometimes be viewed as acyclic sequences. For example, when such sequences are concatenated,
the sequences are considered as acyclic sequences, but the outcome might be a cyclic sequence.
The {\bf \emph{length}} of a sequence $\bm{a}=[a_{0}, a_{1}, ~ \cdots ~, a_{n-1}]$ is $n$.
The length of the sequence $\bm{a}$ will be denoted by $\ell(\bm{a})$.
The {\bf \emph{least period}} of a sequence $\bm{a}=[a_{0}, a_{1}, ~ \cdots ~, a_{n-1}]$ is the least positive integer $\pi$ such that
$a_i = a_{i+\pi}$ for all $0 \leq i \leq n-1$, where subscripts are computed modulo $n$.
The least period of the sequence $\bm{a}$ will be denoted by $\pi(\bm{a})$.
An integer $p$ such that $a_i = a_{i+p}$ for all $0 \leq i \leq n-1$ is a {\bf \emph{period}} of the sequence.
This implies that if a sequence $\bm{a}$ has length $k \pi$ and $\pi =\pi(\bm{a})$, then $\kappa \pi$ is a period of $\bm{a}$ if
and only if $\kappa$ divides $k$.

Similar definitions are adopted and given for arrays. In particular, the horizontal period of an $r_1 \times r_2$ array $\cA$
is the least integer $\pi$ such that $\pi$ is a period of each row of $\cA$.
Similarly, the vertical period of an $r_1 \times r_2$ array $\cA$
is the least integer $\pi$ such that $\pi$ is a period of each column of $\cA$.
In all the $r_1 \times r_2$ arrays considered here, the horizontal period will be $r_2$ and the vertical period will be $r_1$.

Let $\bm{a}=a_{0}, a_{1}, \dots$ and $\bm{b}=b_{0}, b_{1}, \dots$ be two sequences (either cyclic or acyclic) of the same length.
the {\bf \emph{(bitwise) addition}} of $\bm{a}$ and $\bm{b}$ is defined as
$\bm{a} + \bm{b} = a_{0} + b_{0}, a_{1} + b_{1}, \dots$ and
the {\bf \emph{bitwise product}} of $\bm{a}$ and $\bm{b}$ is defined as $\bm{a} \cdot \bm{b} = a_{0}b_{0}, a_{1}b_{1}, \dots$.
Furthermore, the bitwise product of two matrices $A=(a_{i,j})$ and $B=(b_{i,j})$ in $\F_2^{r_1 \times r_2}$ is defined
as $A \cdot B = (a_{i,j}b_{i,j})$, where $0 \leq i \leq r-1$ and $0 \leq j \leq t-1$. Similarly, the (bitwise) addition of two matrices
is defined.

For two positive integers $x$ and $y$, let $\gcd(x,y)$ denote the {\bf \emph{greatest common divisor}} of $x$ and $y$.
Similarly, for two polynomials $f(x)$ and $g(x)$, let $\gcd(f(x),g(x))$ denote the {\bf \emph{greatest common divisor}} of $f(x)$ and $g(x)$.
For two positive integers $x$ and $y$, let $\ell cm(x,y)$ denote the {\bf \emph{least common multiple}} of $x$ and $y$.
When two sequences $\bm{a}$ and $\bm{b}$, where $\ell (\bm{a})=\pi (\bm{a})$ and $\ell (\bm{b})=\pi (\bm{b})$ are not of the same period,
the bitwise product $\bm{a} \cdot \bm{b}$ is defined
as the bitwise product of the sequences $(\overbrace{\bm{a}  \bm{a}   ~ \cdots ~  \bm{a}}^{\mu_1 ~ \text{copies~of}~ \bm{a}})$ and
$(\overbrace{\bm{b}  \bm{b}  ~ \cdots ~  \bm{b}}^{\mu_2 ~ \text{copies~of}~ \bm{b}})$, where
$\mu_1 = \frac{\ell cm (\pi (\bm{a}),\pi (\bm{b}))}{\pi (\bm{a})}$ and
$\mu_2 = \frac{\ell cm (\pi (\bm{a}),\pi (\bm{b}))}{\pi (\bm{b})}$. This implies that the sequence $\bm{a} \cdot \bm{b}$
has length and period $\ell cm (\pi (\bm{a}),\pi (\bm{b}))$.
In the same way we define the bitwise addition of $\bm{a} + \bm{b}$.

The following results appear in Chapter 2 of~\cite{Etz24} or can be inferred from the discussion in this chapter.
The first result is a simple observation, but its consequences are of considerable importance throughout the paper.
\begin{proposition}
\label{thm:sh+add}
If $\bm{a}$ and $\bm{b}$ are two sequences generated by the polynomial $f(x)$, then the sequence
$\bm{a} + \bm{b}$ is also generated by the polynomial $f(x)$.
\end{proposition}
\begin{IEEEproof}
The claim follows immediately from the fact that if $\bm{a}$ and $\bm{b}$ satisfy the same linear recurrence,
then $\bm{a} + \bm{b}$ also satisfies the same recurrence.
\end{IEEEproof}

The property implied by Propsition~\ref{thm:sh+add} is called {\bf \emph{the shift-and-add property}} for sequences generated
by the same LFSR, i.e., the same polynomial.

\begin{corollary}
\label{cor:closed_addition}
The sequences generated by a polynomial $f(x)$ are closed under bitwise addition.
\end{corollary}

\begin{definition}
\label{def:exponent}
The {\bf \emph{exponent}} $e(h(x))$ of a polynomial $h(x)$, where $h(0) \neq 0$, is the least positive integer~$e$ such that
$h(x)$ divides $x^e -1$.
\end{definition}

Let $\gamma(x)$ be the polynomial defined by
\begin{equation*}
\label{eq:gammFmatrix}
\gamma(x) = \sum_{i=1}^n c_i x^i (a_{-i} x^{-i} + a_{-i+1} x^{-i+1}+ \cdots + a_{-1} x^{-1} ) = \sum_{i=0}^{n-1} \gamma_i x^i .
\end{equation*}

\begin{proposition}
\label{thm:period=exponent}
If $( a_k )$ is a nonzero sequence generated by the characteristic polynomial $c(x)$
such that ${\gcd (\gamma(x),c(x))=1}$, then $\pi (( a_k )) = e(c(x))$.
\end{proposition}


\begin{corollary}
\label{cor:period=same_e}
If $c(x)$ is an irreducible polynomial, then the least period of its associated sequence $( a_k )$
is the same for each initial state, except for the all-zero state.
\end{corollary}

\begin{corollary}
\label{cor:sameP}
If $c(x)$ is an irreducible polynomial of degree~$n$, then all the nonzero sequences that it generates have
the same period, which is a factor of~$2^n-1$.
\end{corollary}

\begin{definition}
\label{def:primitive1}
An irreducible polynomial of degree $n$ which is a characteristic polynomial of an $\textup{LFSR}_n$ that generates
a sequence of period~${2^n-1}$, is called a {\bf \emph{primitive polynomial}}. The sequence of period $2^n-1$ which it
generates is called a span $n$ {\bf \emph{\textup{M}-sequence}}.
\end{definition}

\begin{definition}
A {\bf \emph{zero factor}} ZF$(n,k)$ with {\bf \emph{exponent}} $k$ is a set of $d$ cyclic sequences
of least period~$k$, which contains each nonzero $n$-tuple exactly once as a window in one of the cycles.
\end{definition}

For a zero factor with $d$ cycles of length $k$, we must have $d \cdot k = 2^n -1$ and $n < k \leq 2^n -1$.
All known parameters $n$, $d$, and $k$, can be inferred from the following result
proved in~\cite{Etz24,Gol67}.

\begin{proposition}
\label{thm:zero_prod_irreduc}
Let $f_i (x)$, $1 \leq i \leq t$, be $t$ distinct irreducible polynomials of degree $n$ and exponent~$e$.
Then the feedback shift register which has the characteristic polynomial $\prod_{i=1}^t f_i (x)$
produces a zero factor with exponent $e$.
\end{proposition}

\begin{definition}
A reducible polynomial has {\bf \emph{uniform exponent}} $e$ if it is a product of distinct irreducible polynomials of exponent~$e$, i.e.,
all the nonzero sequences that it generates have the same period $e$.
\end{definition}

All zero factors whose sequences are generated by polynomials are induced from Proposition~\ref{thm:zero_prod_irreduc}.
The following proposition and its corollary were proved in~\cite[p. 85, Theorem 3.5]{LiNi97}.
\begin{proposition}
\label{thm:unique_expo}
The degree $d$ of an irreducible polynomial with exponent $e$ is equal to the multiplicative order of 2 modulo $e$, i.e.,
$d$ is the smallest integer such that $2^d \equiv 1 ~ (\mmod ~ e)$.
\end{proposition}

\begin{corollary}
\label{cor:unique_expo}
All the irreducible polynomials with exponent $e$ have the same degree $n$. All reducible polynomials
with uniform exponent $e$ are products of irreducible polynomials of the same degree $n$ and exponent $e$.
\end{corollary}

All the reducible polynomials which are discussed in the paper have uniform exponent so when we use exponent for them we mean uniform exponent.

\begin{definition}
For a polynomial $f(x)$, let $S(f(x))$ denote the set of cyclic sequences whose characteristic polynomial is $f(x)$.
The polynomial $f(x)$ {\bf \emph{generates}} these sequences and
each sequence $\bm{a}$ in $S(f(x))$ is {\bf \emph{generated}} by $f(x)$.
\end{definition}

\begin{corollary}
Let $f_i (x)$, $1 \leq i \leq t$, be $t$ distinct irreducible polynomials of degree $n$ and exponent $e$.
The polynomial $\prod_{i=1}^t f_i (x)$ generates all the sequences which are generated by the polynomial $\prod_{i=1}^{t-1} f_i (x)$
Each $((t-1)n)$-tuple is contained as a window in exactly one sequence generated by $\prod_{i=1}^{t-1} f_i (x)$.
Each $(tn)$-tuple is contained as a window in exactly one sequence generated by $\prod_{i=1}^t f_i (x)$.
\end{corollary}

Sometimes it will be important to distinguish between primitive polynomials and irreducible polynomials which are not primitive,
and hence such an irreducible polynomial will be called an {\bf \emph{irreducible non-primitive polynomial}} (in short, {\bf \emph{INP polynomial}}).

Let $c(x)$ be an irreducible polynomial. Consider the irreducible polynomial
\begin{equation}
\label{eq:primitiveP}
\hat{c}(x) \triangleq x^n c(x^{-1})= x^n - \sum_{i=1}^n c_i x^{n-i},
\end{equation}
called the {\bf \emph{companion polynomial}}, and the {\bf \emph{companion matrix}} $C$ of $c(x)$ defined by
$$
C = \left[
\begin{array}{ccccccc}
0 && 0 & \cdots & 0 && c_n \\
1 && 0 & \cdots & 0 && c_{n-1} \\
0 && 1 & \cdots & 0 && c_{n-2} \\
\vdots && \vdots & \ddots & \vdots && \vdots \\
0 && 0 & \cdots & 1 && c_1 \\
\end{array} \right] .
$$

If $\beta$ is a root of $\hat{c} (x)$ whose order is $t$, then it can be proved in a similar way to~\cite[pp. 66--67]{Etz24}
that we can order the powers of $\beta$, using a column-vector representation of a finite field with $\beta^i$ for $0 \leq i < n$ as a basis, by
$$
(\beta^0),(\beta^1),(\beta^2), \ldots, (\beta^{t-1}) ~.
$$
Similarly, a nonzero sequence generated by $c(x)$ has least period $t$.
Let
$$
X_0,X_1,X_2,\ldots,X_{t-1}
$$
be the consecutive $n$-tuples of such a nonzero sequence generated by $c(x)$.
\begin{lemma}
For each $i$, $0 \leq i \leq t-1$, we have $(\beta^{i+1}) = C(\beta^i)=C^{i+1} (\beta^0)$ and $X_{i+1}=X_i C = X_0 C^{i+1}$,
where superscripts and subscripts are taken modulo $t$.
If $b_i$, $0 \leq i \leq t-1$ is an arbitrary set of binary coefficients, then
$$
\sum_{i=0}^{t-1} b_i (\beta^i) = 0
$$
if and only if
$$
\sum_{i=0}^{t-1} b_i X_i = 0 ~.
$$
\end{lemma}

\section{Folding a Sequence into an Array}
\label{sec:folding}

In this section we will describe the folding technique defined by MacWilliams and Sloane~\cite{McSl76}.
The technique was used in~\cite{McSl76} to fold an M-sequence into an array and if the array satisfies certain conditions, then the
outcome is a PRA. We will generalize the proof and show that the conditions can be weakened and if satisfied then the obtained
array is a PRA. In Section~\ref{sec:foldIrr} it will be proved that the same conditions suffice to show that a set of
arrays forms a PRAC. A successful folding of a sequence into an array depends on
the Chinese Remainder Theorem~\cite[pp. 11--13]{Etz24}m which we state as follows:

\begin{theorem}
\label{thm:RemainderTHM}
Let $m_1,m_2$ be two relatively prime positive integers greater than 1, and let $m = m_1 \cdot m_2$.
If~~$0 \leq k < m$, then the system of equations
\begin{equation*}
\label{eq:china2}
\begin{array}{c}
k \equiv i ~ (\mmod ~ m_1) \\
k \equiv j ~ (\mmod ~ m_2)
\end{array}
\end{equation*}
has a unique solution, for any $i$ and $j$ with $0 \leq i < m_1$, $0 \leq j < m_2$.
\end{theorem}
Another result from number theory required for the proof of the construction is
the following consequence of the extended Euclidean algorithm~\cite[pp. 7--9]{Etz24}.
\begin{lemma}
\label{lem:bezout}
If $a$ and $b$ are two positive integers such that $k=\gcd(a,b)$, then there exist two integers $x$ and $y$ such that $k=ax+by$.
\end{lemma}

The {\bf \emph{folding}} technique which will be used to construct an $(r_1,r_2;n_1,n_2)$-PRA in this section
(and also an $(r_1,r_2;n_1,n_2)$-PRAC in the other sections)
is based on folding nonzero sequences
generated by a polynomial which generates a zero factor with exponent $r_1 r_2$ into $r_1 \times r_2$ arrays.
Assume that $\eta=2^{n_1n_2}-1$, $r_1 =2^{n_1}-1$, and $r_2 = \frac{\eta}{r_1}$, where ${\gcd(r_1,r_2)=1}$.
Let $\cS=[s_0s_1s_2 \cdots s_{\eta -1}]$ be a span $n_1n_2$ M-sequence. Write the consecutive
elements of $\cS$ down the southeast diagonals of an $r_1 \times r_2$ array $\cB=\{ b_{ij}\}$,
$0 \leq i \leq r_1 -1$, $0 \leq j \leq r_2 -1$, starting
at $b_{00} , b_{11} , b_{22}$ and so on, where the last position is $b_{r_1 -1,r_2 -1}$.
After writing position~$b_{ij}$ we continue to write $b_{i+1,j+1}$, where $i+1$ is taken modulo $r_1$ and $j+1$
is taken modulo~$r_2$. In other words, $s_k$ will be written in position~$b_{ij}$,
where $i \equiv k ~ (\mmod ~ r_1)$ and $j \equiv k ~ (\mmod ~ r_2)$. By Theorem~\ref{thm:RemainderTHM} each $k$,
where $0 \leq k \leq r_1 r_2-1$, has a different solution for these two equations
and so the diagonal covers the entire array.
Another type of folding to generate pseudo-random arrays was presented
in~\cite{Spa63}. A generalization of the technique by folding sequences into various shapes was considered and analyzed in~\cite{Etz11}.
The following theorem was proved in~\cite{McSl76}.

\begin{theorem}
\label{thm:PRA_MS}
Each $n_1 \times n_2$ nonzero matrix is contained exactly once as a window in the $r_1 \times r_2$ array~$\cB$
obtained by folding a span $n_1 n_2$ M-sequence, where $r_1 = 2^{n_1}-1$
and $r_2 =(2^{n_1 n_2}-1)/{r_1}$. Moreover, $\cB$~has the shift-and-add property, that is, $\cB$ is an $(r_1,r_2;n_1,n_2)$-PRA.
\end{theorem}

The proof of Theorem~\ref{thm:PRA_MS} presented in~\cite{McSl76}
is based on the observation that it is sufficient to prove that in the
upper-left $n_1 \times n_2$ window of the array $\cB$ we cannot have the all-zero $n_1 \times n_2$ matrix.

If the length of $\cS$ is $r_1 r_2$, where $\gcd(r_1,r_2)=1$ we denote by $\cF(\cS;r_1,r_2)$ the matrix $\cB$ of size $r_1 \times r_2$ obtained
by folding $\cS$ into $\cB$.

\begin{example}
\label{ex:foldPR}
For $n_1=n_2=2$, $r_1=3$, $r_2=5$,
consider the span $4$ ${\textup{M}\text{-sequence}}$ $\cS=[000111101011001]$, with positions numbered
from $0,1$, up to~$14$. Consider now the $3 \times 5$ array $\cB$ with the entries~$b_{ij}$,
$0 \leq i \leq 2$, $0 \leq j \leq 4$, where the positions $0$ through $14$, of the sequence, are folded into $\cB$ as follows
$$
\left[
\begin{array}{ccccc}
b_{00} & b_{01} & b_{02} & b_{03} & b_{04} \\
b_{10} & b_{11} & b_{12} & b_{13} & b_{14} \\
b_{20} & b_{21} & b_{22} & b_{23} & b_{24}
\end{array}
\right],~
\left[
\begin{array}{ccccc}
0 & 6 & 12 & 3 & 9 \\
10 & 1 & 7 & 13 & 4 \\
5 & 11 & 2 & 8 & 14
\end{array}
\right] .
$$
The $\textup{M}$-sequence $\cS$ is folded into the array $\cB$ to produce the array,
$$
\cF(\cS;3,5)= \left[
\begin{array}{ccccc}
0 & 1 & 0 & 1 & 0 \\
1 & 0 & 0 & 0 & 1 \\
1 & 1 & 0 & 1 & 1
\end{array}
\right]
$$
which forms a $(3,5;2,2)$-$\textup{PRA}$.
\hfill\quad $\blacksquare $
\end{example}

If we shift the array $\cB$ horizontally and/or vertically, we obtain an array that is a folding of a shift of $\cS$.
Since the sequences obtained from any primitive polynomial have the shift-and-add property, it follows that if we add
$\cB$ to one of its non-constant shifts, the outcome is another non-constant shift of the array~$\cB$.
This property is the shift-and-add property of the array. This property is also derived by observing that the folding operation
preserves the operation of bitwise addition of sequences. A similar property is satisfied for the bitwise product.
This is summarized in the following lemma.

\begin{lemma}
\label{lem:preserveA_M}
If $\bm{u}$ and $\bm{v}$ are sequences of length $r_1 r_2$ such that $\gcd(r_1,r_2)=1$ then
\begin{enumerate}
\item $\cF(\bm{u};r_1,r_2)+\cF(\bm{v};r_1,r_2)=\cF(\bm{u}+\bm{v};r_1,r_2)$;
\item $\cF(\bm{u};r_1,r_2) \cdot \cF(\bm{v};r_1,r_2)=\cF(\bm{u} \cdot \bm{v};r_1,r_2)$.
\end{enumerate}
\end{lemma}

\begin{example}
\label{ex:fold_PM}
Consider the array and the $\textup{M}$-sequence~$\cS$ of Example~\ref{ex:foldPR}. We shift the array horizontally by $2$ and vertically
by $1$ and add them as follows, where the first bit of $\cS$ is in bold.
$$
\left[
\begin{array}{ccccc}
{\bf 0} & 1 & 0 & 1 & 0 \\
1 & 0 & 0 & 0 & 1 \\
1 & 1 & 0 & 1 & 1
\end{array}
\right] +
\left[
\begin{array}{ccccc}
1 & 1 & 1 & 1 & 0 \\
1 & 0 & {\bf 0} & 1 & 0 \\
0 & 1 & 1 & 0 & 0
\end{array}
\right]
=\left[
\begin{array}{ccccc}
1 & 0 & 1 & 0 & {\bf 0} \\
0 & 0 & 0 & 1 & 1 \\
1 & 0 & 1 & 1 & 1
\end{array}
\right] .
$$
The \textup{M}-sequence $\cS$ starts in the leftmost array in $b_{00}$, in the middle array at $b_{12}$, and in the rightmost array at $b_{04}$.
\hfill\quad $\blacksquare $
\end{example}

\begin{definition}
A {\bf \emph{constant polynomial}} is a polynomial whose degree is zero (note that also the zero polynomial is a constant polynomial).
The {\bf \emph{minimal polynomial}} $m(x)$ of an element $\alpha$
in a finite field $\F_{2^n}$ is the monic polynomial over $\F_{2^m}$ of the least degree such that $m(\alpha)=0$, where $\F_{2^m}$
is a subfield of $\F_{2^n}$.
\end{definition}

Theorem~\ref{thm:PRA_MS} is generalized to obtain more PRAs, with our first main technique, as follows.
\begin{theorem}
\label{thm:HuiminGen}
Let $n_1$, $n_2$, $r_1$, and $r_2$, be positive integers such that $r_1 r_2 = 2^{n_1 n_2}-1$ and $\gcd(r_1,r_2)=1$.
If $r_1$ divides $2^{n_1}-1$ and the $n_1$ integers, $2^i$, $0 \leq i \leq n_1 -1$, are distinct modulo $r_1$, then
folding an \textup{M}-sequence whose length is $r_1 r_2$
into an $r_1 \times r_2$ array $\cB$ forms an $(r_1,r_2;n_1,n_2)$-PRA.
\end{theorem}
\begin{IEEEproof}
Let $\alpha$ be a primitive element in $\F_{2^{n_1 n_2}}$ and let $\cS$ be its associated M-sequence. Furthermore, let $G$ be
an $(n_1 n_2) \times (2^{n_1n_2}-1)$ matrix defined by the $n_1 n_2$ consecutive shifts of the M-sequence $\cS$. In such representation,
the columns of $G$ are arranged in the order isomorphic to $\alpha^0,\alpha^1,\alpha^2,\ldots,\alpha^{2^{n_1n_2}-2}$ defined by the companion matrix.
Let $\ell(i,j)$ be the integer with $0 \leq \ell(i,j) \leq 2^{n_1n_2}-1$ such that
$\ell (i,j) \equiv i ~ (\mmod ~ r_1)$ and $\ell (i,j) \equiv j ~ (\mmod ~ r_2)$.

We claim that the $n_1n_2$ columns of $G$
associated with the elements $\alpha^{\ell(i,j)}$, where
$0 \leq i \leq n_1 -1$
and ${0 \leq j \leq n_2 -1}$, are linearly independent. These columns are associated with the $n_1 \times n_2$ sub-array
in the upper-left corner of $\cB$.

Assume for the contrary that
\begin{equation*}
\sum_{i=0}^{n_1-1} \sum_{j=0}^{n_2-1} c_{ij} \alpha^{\ell(i,j)} =0,
\end{equation*}
where not all the $c_{ij}$ are \emph{zero}. Since $\gcd(r_1,r_2)=1$, it follows by Lemma~\ref{lem:bezout}
that there exist two integers $\mu$ and $\nu$ such that $\mu r_1 + \nu r_2 =1$. Let $\beta = \alpha^{\nu r_2}$,
$\gamma=\alpha^{\mu r_1}$, which implies that $\beta$ is an element of order~$r_1$.
To see that, assume for a contradiction that $\beta^t =1$ for $t < r_1$, i.e.,
$\beta^t =\alpha^{\nu r_2 t} =1$. The order of $\alpha$ is $r_1 r_2$ and hence $r_1r_2$ divides $\nu r_2 t$,
i.e., $r_1$ divides $\nu t$. Since $\mu r_1 + \nu r_2 =1$ we have that $\gcd(r_1,\nu)=1$ which implies that $r_1$ divides $t$,
i.e., $r_1 \leq t$, a contradiction. Similarly, $\gamma$ is an element
of order $r_2$. This also implies that $\alpha = \alpha^{\mu r_1 + \nu r_2}=\beta \gamma$ and hence
$\alpha^{\ell(i,j)} = \beta^{\ell(i,j)} \gamma^{\ell(i,j)}=\beta^i \gamma^j$. Therefore, we have
\begin{equation}
\label{eq:for 2D}
0=\sum_{i=0}^{n_1-1} \sum_{j=0}^{n_2-1} c_{ij} \alpha^{\ell(i,j)}
= \sum_{i=0}^{n_1-1} \sum_{j=0}^{n_2-1} c_{ij} (\beta \gamma)^{\ell(i,j)}
= \sum_{j=0}^{n_2-1} \left( \sum_{i=0}^{n_1-1} c_{ij} \beta^i \right) \gamma^j .
\end{equation}
We claim that $\sum_{i=0}^{n_1-1} c_{i,j} \beta^i =0$.

Since $\beta^{r_1}=1$ and $r_1$ divides $2^{n_1} -1$,
we have that $\beta \in \F_{2^{n_1}}$ and hence $\sum_{i=0}^{n_1-1} (c_{i,j} \beta^i) \in \F_{2^{n_1}}$, i.e., the
coefficient of $\gamma^j$ in Equation~(\ref{eq:for 2D}) is an element of $\F_{2^{n_1}}$.

Let $m$ be the smallest positive integer such that $\gamma^{2^m-1}=1$.
Clearly, $\gamma^{2^m -1} =\alpha^{\mu r_1 (2^m-1)}=1$ and since the order of $\alpha$ is $r_1 r_2$, it follows that
$r_1 r_2$ divides $\mu r_1 (2^m-1)$ and hence $r_2$ divides $\mu (2^m -1)$. Moreover, since $\mu r_1 + \nu r_2 =1$,
it follows that $\gcd(\mu,r_2)=1$ and therefore $r_2$ divides $2^m -1$. We
have that $r_2 = \frac{2^{n_1n_2}-1}{r_1} = \frac{2^{n_1n_2}-1}{2^{n_1}-1} d$, where $d=\frac{2^{n_1}-1}{r_1}$.
Since $r_2$ divides $2^m-1$, it follows that $\frac{2^{n_1n_2}-1}{2^{n_1}-1} d$ divides $2^m-1$ and
hence $\frac{2^{n_1n_2}-1}{2^{n_1}-1}$ divides $2^m-1$ and $d$ divides $2^m-1$.

The binary representation of $\frac{2^{n_1n_2} -1}{2^{n_1}-1}$ (which divides $2^m-1$) is
$10^{n_1 -1} 10^{n_1 -1} 1 \cdots 0^{n_1 -1}1$, where the number of \emph{ones} in this representation is~$n_2$.
Hence, this binary representation contains $n_1 (n_2-1) +1$ digits.
The binary representation of $2^m-1$ is $11 \cdots 1$, where the number of \emph{ones} in this representation is $m$.
Hence, by considering binary multiplication, the smallest~$m$ for which $r_2$ divides $2^m-1$ is~$n_1n_2$.
Thus, $m=n_1 n_2$ is the smallest positive integer such that $\gamma^{2^m}=\gamma$,
the $n_2$~elements
$$
\gamma, \gamma^{2^{n_1}}, \gamma^{2^{2n_1}}, \ldots , \gamma^{2^{(n_2 -1)n_1}}
$$
are distinct.
Therefore,
$$
\prod_{i=0}^{n_2-1} (x - \gamma^{2^{in_1}})
$$
is the minimal polynomial of $\gamma$ in $\F_{2^{n_1}}$. This polynomial has degree~$n_2$ in~$\F_{2^{n_1}}$.
Now, the polynomial on the right side of Equation~(\ref{eq:for 2D}) is a polynomial in $\gamma$ with a smaller degree $n_2-1$ which
is equal to \emph{zero}.

This implies that all the coefficients of $\gamma^j$ in Equation~(\ref{eq:for 2D}) are equal to \emph{zero}.
Therefore, for each $0 \leq j \leq n_2 -1$ we have
$$
\sum_{i=0}^{n_1-1} c_{ij} \beta^i =0 .
$$
This is a contradiction, since this is a polynomial in $\beta$ of degree less than $n_1$ and the minimal polynomial of $\beta$ over~$\F_2$
has degree $n_1$ (since $r_1$ divides $2^{n_1}-1$ we have that $\beta^{2^{n_1}}=\beta$. Moreover, $\beta^{2^0}$, $\beta^{2^1}$,...,$\beta^{2^{n_1 -1}}$,
are distinct and hence $\beta$ is a zero of an irreducible polynomial of degree $n_1$ over $\F_2$ and this is its minimal polynomial.)
This completes the proof of the claim that the $n_1n_2$ columns of $G$ associated with the elements $\alpha^{\ell(i,j)}$,
where $\ell(i,j) \equiv i~(\mmod~r_1)$, $\ell(i,j) \equiv j~(\mmod~r_2)$, $0 \leq i \leq n_1 -1$
and $0 \leq j \leq n_2 -1$, are linearly independent.

This claim implies that the $n_1 \times n_2$ array in the upper-left corner of $\cB$
is nonzero. This $n_1 \times n_2$ window can be chosen arbitrarily since the M-sequence can start at any nonzero initial $(n_1n_2)$-tuple.
This $n_1 \times n_2$ window determines the rest of the elements in $\cB$.
Hence, by the shift-and-add property, there are no two equal $n_1 \times n_2$ windows, as otherwise we can have an
all-zero $n_1 \times n_2$ window by adding the associated two such shifts with two equal $n_1 \times n_2$ windows.
Thus, we have the window property.
\end{IEEEproof}

\section{Folding Sequences of a Zero Factor}
\label{sec:foldZ}

Folding one M-sequence is generalized to folding all the (nonzero) cyclic sequences of a zero factor.
Let $\cC$ be a set of sequences of length $r_1 r_2$, such that $\gcd(r_1,r_2)=1$.
We denote by $\cF(\cC;r_1,r_2)$ the set of $r_1 \times r_2$ arrays obtained by folding the sequences in $\cC$ into $r_1 \times r_2$ arrays.

\begin{example}
\label{ex:fold_seqs}
Let $r_{1}=3$, $r_{2}=7$, $n_{1}=2$, and $n_{2}=3$.
Let $f(x) = x^6 + x^5 + x^4 + x^2 + 1$
be the irreducible polynomial of degree 6 and exponent 21. It generates a ZF$(6,21)$ $\cC$.
The nonzero sequences of $\cC$ generated by $f(x)$ are
\begin{align*}
        & [000001010010011001011], \\
        & [010000111101101010111], \\
        & [001000110111111001110].
\end{align*}
Folding these three sequences of $\cC$ into $3 \times 7$ arrays yields the following three arrays
\begin{align*}
 & \cF(\cC;3,7)= \left\{ \left[\begin{array}{rrrrrrr}
            0 & 0 & 0 & 0 & 0 & 0 & 0 \\
            1 & 0 & 0 & 1 & 0 & 1 & 1 \\
            1 & 0 & 0 & 1 & 0 & 1 & 1
            \end{array}\right], ~~
\left[\begin{array}{rrrrrrr}
            0 & 0 & 1 & 0 & 1 & 1 & 1 \\
            1 & 1 & 1 & 0 & 0 & 1 & 0 \\
            1 & 1 & 0 & 0 & 1 & 0 & 1
            \end{array}\right], ~~
\left[\begin{array}{rrrrrrr}
            0 & 0 & 1 & 0 & 1 & 1 & 1 \\
            1 & 0 & 0 & 1 & 0 & 1 & 1 \\
            1 & 0 & 1 & 1 & 1 & 0 & 0
\end{array}\right] \right\}.
\end{align*}
It is easy to verify that these three arrays form a $(3,7;2,3)$-PRAC.
\hfill\quad $\blacksquare$
\end{example}

It is readily verified that each row of the arrays in Example~\ref{ex:fold_seqs} is either the M-sequence $[0010111]$ or the
all-zero sequence. Each column is either the M-sequence $[011]$ or the all-zero sequence. This phenomena will be explained
in Section~\ref{sec:irreducible}. Meanwhile, the following lemma used to prove this property is easily verified.
\begin{lemma}
\label{lem:prop_m}
Let $r_1$ and $r_2$ be two positive integers such that $\gcd(r_1,r_2)=1$ and
let $\cS = [\bm{a} \,  \bm{a} \,  \dots \,  \, \bm{a}]$ be a sequence of length $r_1 r_2$.
\begin{enumerate}
\item[(1)] If the length of $\bm{a}$ is $r_2$, then each row of $\cF(\cS;r_1,r_2)$ is given by $\bm{a}$ and each column
is either the all-one sequence or the all-zero sequence.

\item[(2)] If the length of $\bm{a}$ is $r_1$, then each column of $\cF(\cS;r_1,r_2)$ is given by $\bm{a}$ and each row
is either the all-one sequence or the all-zero sequence.
\end{enumerate}
\end{lemma}
\begin{IEEEproof}
Part (1) follows from the fact the length of a row in $\cF(\cS;r_1,r_2)$ is the length of the sequence $\bm{a}$ and the consecutive bits of the sequence
are written column by column starting from the first column. Therefore, the $i$-th bit of the sequence $\bm{a}$ is always written to
the $i$-th column of $\cF(\cS;r_1,r_2)$.

Part (2) follows from the fact the length of a column in $F(\cS;r_1,r_2)$ is the length of the sequence $\bm{a}$ and the consecutive
bits of the sequence
are written row by row starting from the first row. Therefore, the $i$-th bit of the sequence $\bm{a}$ is always written to
the $i$-th row of $\cF(\cS;r_1,r_2)$.
\end{IEEEproof}

\begin{example}
\label{ex:folding_columns}
If $\bm{a}=[011]$ and $\cS = [011 \, 011 \,011 \, 011 \, 011 \, 011 \,  011] \in \F_{2}^{3 \cdot 7}$, then
$$
\cA \triangleq \cF(\cS;3,7) =
\left[\begin{array}{rrrrrrr}
            0 & 0 & 0 & 0 & 0 & 0 & 0 \\
            1 & 1 & 1 & 1 & 1 & 1 & 1 \\
            1 & 1 & 1 & 1 & 1 & 1 & 1
\end{array}\right].
$$
\hfill\quad $\blacksquare$
\end{example}

\begin{example}
\label{ex:folding_rows}
If $\bm{a}=[1001011]$ and $\cS = [1001011 \, 1001011 \, 1001011] \in \F_2^{3 \cdot 7}$, then
$$
\cB \triangleq \cF(\cS;3,7) =
\left[\begin{array}{rrrrrrr}
            1 & 0 & 0 & 1 & 0 & 1 & 1 \\
            1 & 0 & 0 & 1 & 0 & 1 & 1 \\
            1 & 0 & 0 & 1 & 0 & 1 & 1
\end{array}\right].
$$
\hfill\quad $\blacksquare$
\end{example}

\begin{lemma}
\label{lem:prod_of_two}
Let $\cA$ be an $r_1 \times r_2$ array for which each row is given by a sequence $\bm{u}$ and each column
is either the all-one sequence or the all-zero sequence.
Let $\cB$ be an $r_1 \times r_2$ array for which each column is given by a sequence $\bm{v}$ and each row
is either the all-one sequence or the all-zero sequence.
Then, the bitwise product $\cA \cdot \cB$ is an $r_1 \times r_2$ array
for which each row is either an all-zero sequence or the sequence~$\bm{u}$ and
each column of $\cA \cdot \cB$ is either an all-zero sequence or the sequence $\bm{v}$.
\end{lemma}
\begin{IEEEproof}
If the $i$-th row of $\cB$ is an all-zero sequence, then the bitwise product yields an all-zero sequence
and hence the $i$-th row of $\cA \cdot \cB$ is the all-zero sequence.
If the $i$-th row of $\cB$ is an all-one sequence, then the bitwise product with the sequence $\bm{u}$ yields the sequence $\bm{u}$
and hence the $i$-th row of $\cA \cdot \cB$ is the sequence $\bm{u}$.

If the $i$-th column of $\cA$ is an all-zero sequence, then the bitwise product yields an all-zero sequence
and hence the $i$-th column of $\cA \cdot \cB$ is the all-zero sequence.
If the $i$-th column of $\cB$ is an all-one sequence, then the bitwise product with the sequence $\bm{v}$ yields the sequence $\bm{v}$
and hence the $i$-th row of $\cA \cdot \cB$ is the sequence $\bm{v}$.
\end{IEEEproof}

\begin{example}
Let $\cA$ and $\cB$ be the $3 \times 7$ arrays of Example~\ref{ex:folding_columns} and~\ref{ex:folding_rows}, respectively.
The bitwise product of $\cA$ and $\cB$, $\cA \cdot\cB$, is as follows
$$
\cA \cdot \cB =
\left[\begin{array}{rrrrrrr}
            0 & 0 & 0 & 0 & 0 & 0 & 0 \\
            1 & 1 & 1 & 1 & 1 & 1 & 1 \\
            1 & 1 & 1 & 1 & 1 & 1 & 1
\end{array}\right]
\cdot
\left[\begin{array}{rrrrrrr}
            1 & 0 & 0 & 1 & 0 & 1 & 1 \\
            1 & 0 & 0 & 1 & 0 & 1 & 1 \\
            1 & 0 & 0 & 1 & 0 & 1 & 1
\end{array}\right] =
\left[\begin{array}{rrrrrrr}
            0 & 0 & 0 & 0 & 0 & 0 & 0 \\
            1 & 0 & 0 & 1 & 0 & 1 & 1 \\
            1 & 0 & 0 & 1 & 0 & 1 & 1
\end{array}\right]
$$
\hfill\quad $\blacksquare$
\end{example}

For the rest of this section
let $f_1(x)$ be a polynomial which forms a zero factor with exponent $r_1$ and $f_2(x)$ be a polynomial
which forms a zero factor with exponent $r_2$.
\begin{lemma}
\label{lem:add2Mat}
If $\cA$ and $\cB$ are two distinct $r_1 \times r_2$ arrays whose columns are either all-zero sequences
or a sequences of length $r_1$ contained in $S(f_1(x))$ and whose rows are either all-zero sequences
or sequences of length $r_2$ contained in $S(f_2(x))$,
then the addition $\cA+\cB$ is also an $r_1 \times r_2$ array
whose columns are either all-zero sequences or sequences contained in $S(f_1(x))$ and
whose rows are either all-zero sequences or a sequences contained in $S(f_2(x))$.
\end{lemma}
\begin{IEEEproof}
Since the columns of $\cA$ and $\cB$ (including those which are all-zero sequences) are sequences generated by $f_1(x)$, it follows by
Proposition~\ref{thm:sh+add} that their addition is also a sequence generated by $f_1(x)$. Hence, each column of
$\cA + \cB$ is either an all-zero sequence or a sequence contained in $S(f_1(x))$. Similarly, a row of $\cA+\cB$ is
either an all-zero sequence or a sequence contained in $S(f_2(x))$.
\end{IEEEproof}

\begin{example}
Consider the primitive polynomial polynomials $f_1(x)= x^2 + x +1$ whose zero factor has exactly one sequence $[011]$
and the reducible polynomial $f_2 (x) = x^6 + x^5 + x^4 + x^3 + x^2 + x+1=(x^3 + x^2 +1)(x^3+x+1)$ whose zero factor $\cC$ has
nine sequences $[0011101]$, $[0010111]$, $[0000101]$, $[0101011]$, $[0001111]$, $[0111111]$, $[0001001]$, $[0000011]$, $[0011011]$.
In the following $3 \times 7$ array~$\cA$, each column is either all-zero sequence or the sequence $[011]$ (or a shift of it). Each row is either the
all-zero sequence or the sequence $[0001111]$ (or a shift of it).
$$
\cA \triangleq
\left[\begin{array}{rrrrrrr}
            0 & 0 & 0 & 0 & 0 & 0 & 0 \\
            0 & 0 & 0 & 1 & 1 & 1 & 1 \\
            0 & 0 & 0 & 1 & 1 & 1 & 1
\end{array}\right] .
$$
Adding $\cA$ to one of its shifts $\cB$ we obtain the following array $\cT_1$:
$$
\cT_1 \triangleq \cA + \cB =
\left[\begin{array}{rrrrrrr}
            0 & 0 & 0 & 0 & 0 & 0 & 0 \\
            0 & 0 & 0 & 1 & 1 & 1 & 1 \\
            0 & 0 & 0 & 1 & 1 & 1 & 1
\end{array}\right]
+
\left[\begin{array}{rrrrrrr}
            0 & 0 & 1 & 1 & 1 & 1 & 0 \\
            0 & 0 & 1 & 1 & 1 & 1 & 0 \\
            0 & 0 & 0 & 0 & 0 & 0 & 0
\end{array}\right] =
\left[\begin{array}{rrrrrrr}
            0 & 0 & 1 & 1 & 1 & 1 & 0 \\
            0 & 0 & 1 & 0 & 0 & 0 & 1 \\
            0 & 0 & 0 & 1 & 1 & 1 & 1
\end{array}\right] .
$$
Adding $\cT_1$ to one of its shifts $\cT_2$ we obtain the following array:
$$
\cT_1 + \cT_2 =
\left[\begin{array}{rrrrrrr}
            0 & 0 & 1 & 1 & 1 & 1 & 0 \\
            0 & 0 & 1 & 0 & 0 & 0 & 1 \\
            0 & 0 & 0 & 1 & 1 & 1 & 1
\end{array}\right]
+
\left[\begin{array}{rrrrrrr}
            0 & 1 & 1 & 1 & 1 & 0 & 0 \\
            1 & 1 & 1 & 1 & 0 & 0 & 0 \\
            1 & 0 & 0 & 0 & 1 & 0 & 0
\end{array}\right] =
\left[\begin{array}{rrrrrrr}
            0 & 1 & 0 & 0 & 0 & 1 & 0 \\
            1 & 1 & 0 & 1 & 0 & 0 & 1 \\
            1 & 0 & 0 & 1 & 0 & 1 & 1
\end{array}\right] .
$$
Each column of $\cA + \cB$ and of $\cT_1 + \cT_2$ is either the all-zero sequence or the sequence $[011]$
and each row is either the all-zero sequence or a sequence from $\cC$.
\hfill\quad $\blacksquare$
\end{example}

Let $S(f_1(x))$ and $S(f_2(x))$ be the set of nonzero sequences
generated by $f_1(x)$ and $f_2(x)$, respectively (these sequences are taken in all their shifts).
Assume that all the sequences in $S(f_i(x))$ have least period $r_i$, $i=1,2$ and $\gcd(r_1,r_2)=1$. Let
$$
\C \triangleq \{ \cF(\bm{a} \cdot \bm{b};r_1,r_2) ~:~ \bm{a} \in S(f_1(x)), ~~ \bm{b} \in S(f_2(x)) \}~.
$$

\begin{lemma}
\label{lem:basicArray}
Let $\bm{a}$ be a nonzero sequence generated by $f_1(x)$ and let $\bm{b}$ be a nonzero sequence generated by $f_2(x)$.
If $\cS_1 \triangleq [\bm{a} \, \bm{a} \, \cdots \, \bm{a}] \in \F_2^{r_1 \cdot r_2}$ and
$\cS_2 \triangleq [\bm{b} \, \bm{b} \, \cdots \, \bm{b}] \in \F_2^{r_1 \cdot r_2}$,
then the $r_1 \times r_2$ array $\cF(\cS_1;r_1,r_2) \cdot \cF(\cS_2;r_1,r_2)$ is contained in $\C$.
\end{lemma}
\begin{IEEEproof}
The claim in the lemma follows directly from Lemma~\ref{lem:preserveA_M}.
\end{IEEEproof}

\begin{corollary}
\label{cor:rowINfold}
If $\tilde{\bm{a}}$ is an $n_1$-tuple in the sequence $\bm{a}$ and $\tilde{\bm{b}}$ is an $n_2$-tuple in the sequence $\bm{b}$,
then the $n_1 \times n_2$ matrix $M$ whose $i$-th row is $\tilde{\bm{b}}$ if the $i$-th entry of $\tilde{\bm{a}}$ is \emph{one}
and all-zero if the $i$-th entry of $\tilde{\bm{a}}$ is \emph{zero}, is contained in $\cF(\bm{a} \cdot \bm{b};r_1,r_2)$.
\end{corollary}

\begin{corollary}
\label{cor:columnINfold}
If $\tilde{\bm{a}}$ is an $n_1$-tuple in the sequence $\bm{a}$ and $\tilde{\bm{b}}$ is an $n_2$-tuple in the sequence $\bm{b}$,
then the $n_1 \times n_2$ matrix $M$ whose $i$-th column is $\tilde{\bm{a}}$ if the $i$-th entry of $\tilde{\bm{b}}$ is \emph{one}
and all-zero if the $i$-th entry of $\tilde{\bm{b}}$ is \emph{zero}, is contained in $\cF(\bm{a} \cdot \bm{b};r_1,r_2)$.
\end{corollary}

\begin{definition}
Let $S(f_{1}(x)) \cdot S(f_{2}(x))$ be defined as a vector space
spanned by all products $\bm{a} \cdot \bm{b}$, where $\bm{a} \in S(f_{1}(x))$ and $\bm{b} \in S(f_{2}(x))$.
\end{definition}

For the following lemma we are not aware of a proof in~\cite{LiNi97} although we assume that the authors were aware of the result.
The lemma can be proved by careful enumeration (see the proof in~\cite[pp. 61--63, Lemma 2.5 and Theorem 2.5]{Etz24}).
It is also a corollary of Proposition~\ref{thm:two_polys} below.

\begin{lemma}
\label{lem:period}
Let $S(f_1(x))$ and $S(f_1(x))$ be the set of nonzero sequences
generated by $f_1(x)$ and $f_2(x)$, respectively.
Assume that all the sequences in $S(f_i(x))$ have least period $r_i$, $i=1,2$. If $\gcd(r_1,r_2)=1$, then the least period of
the nonzero sequences in the linear span of the sequences in the set
$$
\C \triangleq  \{ \bm{a} \cdot \bm{b} ~:~ \bm{a} \in S(f_1(x)), ~~ \bm{b} \in S(f_2(x)) \}
$$
is $r_1 r_2$.
\end{lemma}

\section{A Necessary and Sufficient Condition for Irreducible Polynomials}
\label{sec:sufficient}

The proof in~\cite{McSl76} that the array $\cB$ is an $(r_1,r_2;n_1,n_2)$-PRA requires that
$r_1=2^{n_1}-1$. It appears that this requirement is necessary in some cases, but it is not required
in other cases (see Theorem~\ref{thm:HuiminGen}). The proof of Theorem~\ref{thm:PRA_MS} given in~\cite{McSl76}
and also the proof of Theorem~\ref{thm:HuiminGen} is based on the observation that it is sufficient to show that the
$n_1 \times n_2$ all-zero matrix is not contained as a window of the array~$\cB$.
In this section, a necessary and sufficient condition that
an array (or set of arrays) constructed by folding sequences generated by an irreducible polynomial
is a pseudo-random array (or a pseudo-random array code) will be given. This condition is key for constructing such arrays
and array codes, respectively,
but can be inefficient to verify computationally. Later in the paper, more direct (but less general)
conditions will be presented for the families of codes constructed in the paper.
Section~\ref{sec:second_Necc_Suff} presents an alternative, more efficient, general condition.

For this purpose, we develop a simple theory. The concepts of this theory
are similar to the concepts developed for VLSI testing in Lempel and Cohn~\cite{LeCo85}. However, the main proof provided here is different since the
proof in~\cite{LeCo85} does not hold for the generalized theory which will be presented.
In VLSI testing a test sequence $\blds$ is called {\bf $(\blds,t)$\emph{-universal}} if it exercises every function depending on $t$ or
fewer inputs on a VLSI chip with $s$ inputs. The problem has attracted a lot of interest and various solutions were
suggested. Some of these solutions are based on FSRs, e.g.,~\cite{BCR83,Hol90,LeCo85,TaCh84,WaMc88}.

\begin{definition}
For a set $P= \{p_0,p_1,\ldots,p_{t-1} \}$ of $t$ positions in a sequence, the {\bf \emph{set polynomial}} $g_P(x)$
is defined by
$$
g_P(x) \triangleq \prod_{Q \subseteq P} \sum_{p_i \in Q} x^{p_i}~.
$$
\end{definition}

Let $f(x) = 1 +\sum_{i=1}^n c_i x^i$, where $c_n=1$ be irreducible polynomial with exponent $e$.
Let $\cA=[a_0a_1a_2a_3 \cdots a_{e-1}]$ be the set of cyclic sequences with characteristic polynomial $f(x)$, that satisfy
the recurrence in Equation~(\ref{eq:recur_seq}). So $\cA$ consists of a single M-sequence if $f(x)$ is primitive,
and consists of several sequences all of period $e$ otherwise;
see Corollary~\ref{cor:sameP} and Proposition~\ref{thm:zero_prod_irreduc}.
Consider all the possible shifts of the nonzero sequences in $\cA$ (generated by $f(x)$) as rows in a $(2^n-1) \times e$ matrix $B$, and
let $T$ be the $(2^n-1) \times n$ matrix which is formed by a projection of any $n$~columns of $B$.

\begin{lemma}
\label{lem:tupleA}
Every nonzero $n$-tuple appears as a row of the matrix $T$ if and only if the columns of $T$ are linearly independent.
\end{lemma}
\begin{IEEEproof}
Assume first that each $n$-tuple appears as a row in $T$. This immediately implies that the $n$~columns of $T$ are linearly independent.

Assume now that the columns of $T$ are linearly independent.
Since each nonzero $n$-tuple appears as a window exactly once in one of the nonzero sequences generated by $f(x)$, it follows
that every $n$~consecutive columns of $B$ contain each one of the $2^n-1$ nonzero $n$-tuples as a row.
Hence, the first $n$~columns of $B$ contain each non-zero $n$-tuple exactly once.
These column vectors can be used as rows for the generator matrix of the simplex code of length $2^n-1$ and dimension $n$.
Each other column of $B$ can be represented as a linear combination of the first $n$ columns of $B$.
This linear combination is defined by the recursion induced by $f(x)$ given in Equation~(\ref{eq:recur_seq}).
Hence, all these linear combinations coincide with the codewords of the simplex code.
This implies that every $n$ linearly independent columns contain each nonzero $n$-tuple as a row in $T$.
\end{IEEEproof}

\begin{lemma}
\label{lem:pqA}
Let $f(x)$ be an irreducible polynomial.
If $Q$ is a nonempty subset of $P$ and $q(x) = \sum_{p_i \in Q} x^{p_i}$, then $f(x)$~divides $q(x)$
if and only if the columns of $B$ that are associated with the subset~$Q$ sum to \emph{zero}.
\end{lemma}
\begin{IEEEproof}
If the columns in $B$ that are associated with the subset $Q$ sum to \emph{zero}, then one of the columns
is a sum of some of the other columns, that is, this column is a linear combination of the other columns.
This linear combination is induced by the polynomial $f(x)$ and hence $q(\beta)=0$, where $\beta$
is a root of~$f(x)$. Since we also have $f(\beta)=0$ and $f(x)$ is an irreducible polynomial, it follows that $f(x)$~divides $q(x)$.

If $f(x)$ divides $q(x)$, then $f(\alpha)=0$ implies that $q(\alpha)=0$ and hence since the columns of $B$ are defined
by the recursion induced by $f(x)$, it follows that the columns of $B$ associated with the subset $Q$ sum to \emph{zero}.
\end{IEEEproof}

The given proof of Lemma~\ref{lem:pqA} will not work if $f(x)$ is not an irreducible polynomial
since $f(x)$ does not have to divide $q(x)$ when it shares a non-trivial factor with $q(x)$.

\begin{theorem}
\label{thm:px_gRx}
Given an irreducible polynomial $f(x)$ and a set~$P$ of $n$ coordinates in $B$, then the set~$P$ of coordinates in~$B$ contains each nonzero
$n$-tuple if and only if $g_P (x)$ is not divisible by~$f(x)$.
\end{theorem}
\begin{IEEEproof}
Consider the $(2^n-1) \times n$ matrix $T$ projected by the $n$ columns of $B$ associated with the coordinates of $P$.
By Lemma~\ref{lem:tupleA} every nonzero $n$-tuple appears as a row in $T$ if and only if the columns of $T$
are linearly independent. The columns of $T$ are linearly dependent if and only if
a non-empty subset of the columns in $T$ sums to \emph{zero}.

By Lemma~\ref{lem:pqA} we have that $f(x)$ divides the polynomial $\sum_{p_i \in Q} x^{p_i}$,
where $Q$ is a nonempty subset of~$P$, if and only if the associated subset of columns
of~$T$ sums to \emph{zero}.

Since the polynomial $f(x)$ is irreducible, it follows that $f(x)$ divides the set polynomial $g_P(x)$ if and only if
there exists a subset $Q \subseteq P$ such that $f(x)$~divides the factor $\sum_{p_i \in Q} x^{p_i}$
of $g_P(x)$. Hence, by Lemmas~\ref{lem:tupleA} and~\ref{lem:pqA} the proof is completed.
\end{IEEEproof}

Theorem~\ref{thm:px_gRx} is a necessary and sufficient condition and hence
yields a method to verify whether a sequence that is constructed by a concatenation
of all nonzero sequences generated by an irreducible polynomial is $(\blds,t)$-universal.
The same is true for testing all the arrays generated by folding the sequences generated
by an irreducible polynomial. To examine whether a set of arrays formed by folding is
an $(r_1,r_2;n_1,n_2)$-PRAC we have to examine all the positions in the sequences associated with
$n_1 \times n_2$ windows inthe folding. But, since the positions in two such windows differ in a cyclic shifts it follows that
the set polynomials of two different positions differ by a multiplicative of $x^i$, for some $i$,
which implies that exactly one set polynomial should be examined to verify whether the constructed arrays form an $(r_1,r_2;n_1,n_2)$-PRAC.

\begin{example}
Consider the four INP polynomials -- 1011101001111, i.e., $f_1(x)=x^{12} + x^{10} +x^9 +x^8 +x^6 +x^3 +x^2 +x +1$;
1100101101111, i.e., $f_2(x)=x^{12} +x^{11} +x^8 +x^6 +x^5 +x^3 +x^2 +x +1$;
1110001011111, i.e., $f_3(x)=x^{12} +x^{11} +x^{10} +x^6 +x^4 +x^3 +x^2 +x+1$;
1010011011111, i.e., $f_4(x)=x^{12} +x^{10} +x^7 +x^6 +x^4 +x^3 +x^2 +x+1$.
These four polynomials of degree 12 are four of the 24 polynomials whose exponent is 455.
Now, consider the folding of their sequences into $13 \times 35$ arrays.
For each of these polynomials, we ask whether its folding contain an all-zero $4 \times 3$ array, and also whether
its folding contain an all-zero $3 \times 4$ array.

We start by considering the $4 \times 3$ windows. The entries on the upper-left corner of an array are
$(0,0)$, $(0,1)$, $(0,2)$, $(1,0)$, $(1,1)$, $(1,2)$, $(2,0)$, $(2,1)$, $(2,2)$, $(3,0)$, $(3,1)$, and $(3,2)$.
Each entry $(i,j)$ is translated into a position $k$ in the sequence using the pair of equations
$$
\begin{array}{c}
k \equiv i ~ (\mmod ~ 13) \\
k \equiv j ~ (\mmod ~ 35)
\end{array}
$$
The 12 positions for the set polynomial $g_P(x)$ are
$$
P = \{ 0,~1,~2,~105,~106,~143,~210,~247,~248,~351,~352,~353\}.
$$

We check whether each polynomial divides $g_P(x)$, which implies that it does not produce a $(13,35;4,3)$-PRAC, or it does not divide it, which implies that
the folding its sequences yields a $(13,35;4,3)$-PRAC.
\begin{itemize}
\item The polynomial $f_1(x)$ divides the factor $x^{352}+x^{315}+x^{107}+x^{106}+x^{105}+x+1$ of $g_P(x)$ and hence folding its sequences
into $13 \times 35$ arrays does not yield a $(13,35;4,3)$-PRAC.

\item The polynomial $f_2(x)$ divides the factor $x^{352}+x^{351}+x^{211}+x^{210}+x^{106}+x^{105}+x^2+x+1$ of $g_P(x)$ and hence folding its sequences
into $13 \times 35$ arrays does not yield a $(13,35;4,3)$-PRAC.

\item The polynomial $f_3(x)$ does not divide $g_P(x)$ and hence folding its sequences into $13 \times 35$ arrays yields a $(13,35;4,3)$-PRAC.

\item The polynomial $f_4(x)$ does not divide $g_P(x)$ and hence folding its sequences into $13 \times 35$ arrays yields a $(13,35;4,3)$-PRAC.
\end{itemize}

We continue by considering the $3 \times 4$ windows and the corresponding
12 positions for the set polynomial $g_P(x)$. The positions in the upper-left corner are
$(0,0)$, $(0,1)$, $(0,2)$, $(0,3)$, $(1,0)$, $(1,1)$, $(1,2)$, $(1,3)$, $(2,0)$, $(2,1)$, $(2,2)$, and $(2,3)$.
Each position $(i,j)$ is translated into a position $k$ in the sequence using the pair of equations
$$
\begin{array}{c}
k \equiv i ~ (\mmod ~ 13), \\
k \equiv j ~ (\mmod ~ 35).
\end{array}
$$
The 12 positions for the set polynomial $g_P(x)$ are
$$
P= \{ 0,~1,~2,~105,~106,~105,~210,~211,~247,~315,~351,~352\}.
$$

\begin{itemize}
\item The polynomial $f_1(x)$ divides the factor $x^{353}+x^{352}+x^{248}+x^{247}+x^{210}+x^{143}+x$ of $g_P(x)$ and hence folding its sequences
into $13 \times 35$ arrays does not yield a $(13,35;3,4)$-PRAC.

\item The polynomial $f_2(x)$ does not divide $g_P(x)$ and hence folding its sequences into $13 \times 35$ arrays yields a $(13,35;3,4)$-PRAC.

\item The polynomial $f_3(x)$ divides the factor $x^{353}+x^{352}+x^{247}+x^{210}+x^{106}+x^{105}+x^2+x$ of $g_P(x)$ and hence folding its sequences
into $13 \times 35$ arrays does not yield a $(13,35;3,4)$-PRAC.

\item The polynomial $f_4(x)$ does not divide $g_P(x)$ and hence folding its sequences into $13 \times 35$ arrays yields a $(13,35;3,4)$-PRAC.
\end{itemize}

As we see, each one of the four irreducible polynomials behaves differently in constructing these two pseudo-random array codes.
To summarize the consequences we have
\begin{enumerate}
\item Folding the sequences of $f_1(x)$ does not produce a $(13,35;4,3)$-PRAC and does not produce a $(13,35;3,4)$-PRAC.

\item Folding the sequences of $f_2(x)$ does not produce a $(13,35;4,3)$-PRAC and produces a $(13,35;3,4)$-PRAC.

\item Folding the sequences of $f_3(x)$ produces a $(13,35;4,3)$-PRAC and does not produce a $(13,35;3,4)$-PRAC.

\item Folding the sequences of $f_4(x)$ produces a $(13,35;4,3)$-PRAC and produces a $(13,35;3,4)$-PRAC.
\end{enumerate}
\hfill\quad $\blacksquare$
\end{example}

\section{PRACs Obtained by Folding}
\label{sec:irreducible}

In this section, the second main technique of this paper will be presented. This technique yields a construction for PRACs with various parameters.
Infinite families and specific PRACs obtained by the construction will be analysed in the next section.

\begin{definition}
Let $f_1(x)$ and $f_2(x)$ be two non-constant polynomials. We denote by $f_1(x) \vee f_2(x)$ the polynomial
whose roots have the form $\alpha \beta$, where $\alpha$ is a root of $f_1(x)$ and $\beta$ is a root
of $f_2(x)$ in the splitting fields of $f_1(x) f_2(x)$.
\end{definition}

\begin{example}
\label{ex:primitive}
Let $f_{1}(x) = x^{4}+x+1$ and $f_{2}(x)=x^{3}+x+1$.
The splitting field of $f_{1}(x)f_{2}(x)$ is the finite field $\F_{2^{12}}$.
Let $\alpha$ be a primitive element of $\F_{2^{12}}$ with the minimal ploynomial $m(x) = x^{12} + x^7 + x^6 + x^5 + x^3 + x + 1$.
The roots of $f_{1}(x)$ are
$$
\beta_{1} = \alpha^{273}, \beta_{2} =\alpha^{546}, \beta_{3} =\alpha^{1092}, \beta_{4} = \alpha^{2184}.
$$
The roots of $f_2(x)$ are
$$
\gamma_{1} = \alpha^{585}, \gamma_{2} = \alpha^{1170}, \gamma_{3} =  \alpha^{2340}.
$$
Finally, $f_{1}(x) \vee f_{2}(x) = \prod_{i=1}^{4} \prod_{j=1}^{3} (x - \beta_{i}\gamma_{j}) = x^{12} + x^9 + x^5 + x^4 + x^3 + x + 1$.
\hfill\quad $\blacksquare$
\end{example}

The following proposition implied by Theorem 8.67 in~\cite{LiNi97} is a key result for the construction of PRACs.
More general results were also proved in~\cite{Sel66} and~\cite{ZiMi73}.
It shows a relation between vector spaces $S(f_{1}(x) \vee f_{2}(x))$ and $S(f_{1}(x)) \cdot S(f_{2}(x))$.

\begin{proposition}
\label{thm:two_polys}
Let $f_{1}(x)$ and $f_{2}(x)$ be two polynomials of degrees $n_1$ and $n_2$ and no repeated roots, respectively.
Moreover, suppose  $f_1(x)$ and $f_2(x)$ generate zero factors,
ZF$(n_1,r_1)$ and ZF$(n_2,r_2)$, respectively.
Let $\beta_i$, $1 \leq i \leq n_1$ be the distinct roots of $f_1(x)$ and $\gamma_j$, $1 \leq j \leq n_2$ be the distinct roots of $f_2(x)$
in the splitting fields of $f_{1}(x) f_{2}(x)$.
Then
$S(f_{1}(x)) \cdot S(f_2(x)) = S(f_{1}(x) \vee f_2(x))$, where
$$
f_{1}(x) \vee f_{2}(x) = \prod_{i=1}^{n_1} \prod_{j=1}^{n_2} (x - \beta_i \gamma_j).
$$
\end{proposition}

Proposition~\ref{thm:two_polys} implies in other words that
the set of sequences generated by $f_{1}(x) \vee f_{2}(x)$ is the linear span of
the sequences that are obtained as a bitwise product of the sequences generated by $f_{1}(x)$ and $f_{2}(x)$.
Proposition~\ref{thm:two_polys} leads to another construction of PRACs.

\begin{theorem}
\label{thm:main_res}
Let $r_{1}$, $r_{2}$, $n_{1}$, $n_{2}$ be positive integers such that $\gcd(r_1,r_2)=1$,
and let $f_i(x)$, $i \in \{ 1,2\}$, be a non-constant polynomial of degree $n_i$ and uniform exponent $r_i$.
If the feedback shift register with the characteristic polynomial $g(x) \triangleq f_{1}(x) \vee f_{2}(x)$
produces a zero factor ZF$(n_{1}n_{2}, r_{1}r_{2})$, i.e., $g(x)$ has degree $n_1 n_2$ and uniform exponent $r_1 r_2$,
then folding the sequences generated by $g(x)$ into $r_1 \times r_2$ arrays yields
an $(r_{1},r_{2}; n_{1},n_{2})$-PRAC with $(2^{n_{1}n_{2}}-1) / r_{1}r_{2}$ codewords.
\end{theorem}
\begin{IEEEproof}
Let $\cC$ be the zero factor obtained from the nonzero sequences of the
characteristic polynomial $f_{1}(x) \vee f_{2}(x)$.
We claim that $\cF(\cC;r_1,r_2)$ is an $(r_{1},r_{2}; n_{1},n_{2})$-PRAC with $\ell=(2^{n_{1}n_{2}}-1)/r_{1}r_{2}$ codewords.

Let $\cA$ and $\cA^{\prime}$ be two $r_1 \times r_2$ arrays in $\cF(\cC;r_1,r_2)$, where either $\cA^{\prime}$ is a non-constant shift of $\cA$,
or $\cA^{\prime}$ is different from $\cA$. By definition $\cA = \cF(\bm{v};r_1,r_2)$ and $\cA^{\prime} = \cF(\bm{v}^{\prime};r_1,r_2)$
for some sequences $\bm{v}, \bm{v}^{\prime} \in \cC$,
where either $\bm{v}^{\prime}$ is a non-trivial shift of $\bm{v}$, or $\bm{v}^{\prime}$ is diffferent from $\bm{v}$.
Since $\cC$ is a zero factor generated by an LFSR, it follows from Lemma~\ref{lem:preserveA_M} that
$\cA+\cA^{\prime} = \cF(\bm{v};r_1,r_2) + \cF(\bm{v}^{\prime};r_1,r_2) = \cF(\bm{v} + \bm{v}^{\prime};r_1,r_2)$.
Therefore, $\cA+\cA^{\prime}$ is a codeword in $\cF(\cC;r_1,r_2)$. Since each $n_1 n_2$-tuple appears in exactly one window of a sequence in $\cC$
and each sequence of~$\cC$ has least period $r_1 r_2$ it follows that the number of sequences in $\cC$ is $\ell =(2^{n_{1}n_{2}}-1) / r_{1}r_{2}$
and this is also the number of codewords in $\cF(\cC;r_1,r_2)$.

By sliding an $n_{1} \times n_{2}$ window on all positions of the $\ell$ arrays of $\cF(\cC;r_1,r_2)$,
we obtain a set of $2^{n_{1}n_{2}}-1$ matrices of size $n_{1} \times n_{2}$.
It remains to prove that any two such $n_1 \times n_2$ matrices are distinct and the
$n_1 \times n_2$ all-zero matrix is not contained in this set.
Note that by Proposition~\ref{thm:two_polys}, the matrices in $\cF(\cC;r_1,r_2)$ are spanned by the matrices in
$\{ \cF( \bm{a} \cdot \bm{b},r_1,r_2) ~:~ \bm{a} \in S(f_1(x)), ~ \bm{b} \in S(f_2(x)) \}$.
(The fact that the polynomials have an uniform exponent implies their roots are distinct, and so we can apply Proposition~\ref{thm:two_polys}.)
By Lemmas~\ref{lem:prop_m}, \ref{lem:add2Mat}, and~\ref{lem:basicArray} we have that
a codeword $\cA$ in $\cF(\cC;r_1,r_2)$ has the following two properties:
\begin{enumerate}
\item[(a)] Each column of $\cA$ is either all-zero or a nonzero sequence generated by $f_1(x)$.

\item[(b)] Each row of $\cA$ is either all-zero or a nonzero sequence generated by $f_2(x)$.
\end{enumerate}

Let $\cA \in \cF(\cC;r_1,r_2)$, let $X$ be an arbitrary $n_{1} \times n_{2}$ window from $\cA$, and
assume first that $X$ is an all-zero matrix. Since each nonzero sequence generated by $f_2(x)$ does
not have a run of $n_2$ consecutive \emph{zeros}, it follows that all the rows of $\cA$ that contain $X$ are all-zero rows,
i.e., $\cA$ contains $n_1$ consecutive all-zero rows.
Therefore, since each nonzero sequence generated by $f_1(x)$ does
not have a run of $n_1$ consecutive \emph{zeros}, it follows that all the columns of $\cA$ are all-zero columns,
i.e., $\cA$ is an all-zero matrix, a contradiction.
Thus, the $n_1 \times n_2$ all-zero matrix is not contained in a window of $\cA$.

If two $n_1 \times n_2$ windows from $\cA,\cA' \in \cF(\cC;r_1,r_2)$ form the same matrix, then their bitwise addition
in the related shift forms another codeword with an all-zero $n_1 \times n_2$ window, a contradiction.

Thus, the theorem is proved.
\end{IEEEproof}

Recall that Proposition~\ref{thm:zero_prod_irreduc} exhibits polynomials with a given uniform exponent.
Hence, from Lemma~\ref{lem:period} and Theorem \ref{thm:main_res} we have the following consequence.

\begin{corollary}
\label{cor:main_res1}
Let $r_{1}$, $r_{2}$, $n_{1}$, $n_{2}$ be positive integers such that $\gcd(r_{1},r_{2})=1$.
For $i \in \{ 1,2 \}$, let $f_{i}(x)$ be a non-constant polynomial of degree $n_{i}$ and uniform exponent $r_i$.
Then, the sequences generated by $g(x) \triangleq f_{1}(x) \vee f_{2}(x)$ can be
folded into $r_1 \times r_2$ arrays and form an $(r_{1},r_{2}; n_{1}, n_{2})$-PRAC.
\end{corollary}

\begin{example}
\label{ex:red_red_red}
Let $f_1(x) = (x^3 + x^2 +1)(x^3+x+1)=x^6+x^5+x^4+x^3+x^2+x+1$ be a reducible polynomial with uniform exponent 7 and
$f_2(x) = (x^4+x^3+1)(x^4 +x+1)=x^8 +x^7 +x^5 + x^4 +x^3 +x+1$ be a reducible polynomial with uniform exponent 15.
The polynomial
\begin{align*}
f_1(x) \vee f_2(x)= & x^{48} + x^{47} + x^{46} + x^{43} + x^{42} + x^{40} + x^{39} + x^{36} + x^{35} + x^{34}+ x^{33} + x^{32} + x^{31}+x^{28}\\ &
\hspace{-1.3cm} + x^{26}+ x^{24}+x^{22} + x^{20}+x^{17}+x^{16} + x^{15}+x^{14}+x^{13} + x^{12}+ x^9 + x^8 +x^6 + x^5 +x^2 +x + 1\\ &=(x^{12}+x^8 + x^6 +x^5+x^3+x^2+1)(x^{12}+x^9+x^5+x^4+x^3+x+1)\\ & (x^{12}+x^{10}+x^9+x^7+x^6+x^4+1)(x^{12}+x^{11}+x^9+x^8+x^7+x^3+1)
\end{align*}
is a reducible polynomial with uniform exponent $105$ which forms a ZF$(12,105)$.
By Theorem~\ref{thm:main_res} and Corollary~\ref{cor:main_res1}, folding the sequences of this zero factor into $7 \times 15$ arrays
yields a $(7,15;6,8)$-PRAC.
\hfill\quad $\blacksquare$
\end{example}

\section{Folding Sequences from an Irreducible Polynomial}
\label{sec:foldIrr}

The main goal of this section is to generalize Theorem~\ref{thm:HuiminGen} for pseudo-random array codes. In the next section
we compare the constructed codes with the codes which were constructed based on Theorem~\ref{thm:main_res} and Corollary~\ref{cor:main_res1}.

\begin{theorem}
\label{thm:HuiPRAC}
Let $n_1$, $n_2$, $r_1$, $r_2$ be positive integers such that $r_1 r_2$ divides $2^{n_1n_2}-1$ and ${\gcd(r_1,r_2)=1}$.
If $r_1$ divides $2^{n_1}-1$ and the integers $2^i ~(\mmod ~ r_1)$, $0 \leq i \leq n_1 -1$ are distinct, then
folding the sequences generated by an irreducible polynomial $g(x)$, of degree $n_1n_2$ and exponent is $r_1 r_2$ into
$r_1 \times r_2$ arrays forms an $(r_1,r_2;n_1,n_2)$-PRAC $\cB$.
\end{theorem}
\begin{IEEEproof}
Let $\alpha$ be a primitive element in $\F_{2^{n_1 n_2}}$, $\ell = \frac{2^{n_1 n_2} -1}{r_1r_2}$, and $\delta = \alpha^\ell$.
Note that the order of $\delta$ is $\eta=r_1 r_2$. Let $g(x)$ be an irreducible polynomial, for which
the root of its companion polynomial is $\delta$.
Let $\cS$ be a nonzero sequence generated by $g(x)$. Furthermore, let $G$ be
an $(n_1 n_2) \times (r_1 r_2)$ matrix defined by the $n_1 n_2$ consecutive shifts of $\cS$. In such representation,
the columns of $G$ are arranged in the order isomorphic to $\delta^0,\delta^1,\delta^2,\ldots,\delta^{\eta -1}$ defined by the companion matrix.
For $(i,j)$ with $0 \leq i<r_1$ and $0 \leq j<r_2$, let $\ell(i,j)$ be the integer
with $0\leq \ell(i,j)<\eta$ such that $\ell (i,j) \equiv i ~ (\mmod ~ r_1)$ and $\ell (i,j) \equiv j ~ (\mmod ~ r_2)$.

We claim that the $n_1n_2$ columns of $G$ associated with the elements $\delta^{\ell(i,j)}$,
where $\ell(i,j) \equiv i~(\mmod~r_1)$, $\ell(i,j) \equiv j~(\mmod~r_2)$, $0 \leq i \leq n_1 -1$
and ${0 \leq j \leq n_2 -1}$, are linearly independent. These columns are associated with the $n_1 \times n_2$ sub-array
in the upper-left corner of $\cB$.

Assume for the contrary that
\begin{equation*}
\sum_{i=0}^{n_1-1} \sum_{j=0}^{n_2-1} c_{ij} \delta^{\ell(i,j)} =0,
\end{equation*}
where not all the $c_{ij}$ are \emph{zeros}. Since $\gcd(r_1,r_2)=1$, it follows by Lemma~\ref{lem:bezout}
that there exist two integers $\mu$ and $\nu$ such that $\mu r_1 + \nu r_2 =1$. Let $\beta = \alpha^{\nu r_2}$,
$\gamma=\alpha^{\mu r_1}$, which implies that $\alpha = \alpha^{\mu r_1 + \nu r_2}=\beta \gamma$.
Hence, $\delta^{\ell(i,j)}=(\alpha^\ell)^{\ell(i,j)}=(\beta^i \gamma^j)^\ell$.
Therefore, we have
\begin{equation}
\label{eq:for 2Dprac}
0=\sum_{i=0}^{n_1-1} \sum_{j=0}^{n_2-1} c_{ij} \delta^{\ell(i,j)}
= \sum_{i=0}^{n_1-1} \sum_{j=0}^{n_2-1} c_{ij} (\beta^i \gamma^j)^{\ell}
= \sum_{j=0}^{n_2-1} \left( \sum_{i=0}^{n_1-1} c_{ij} (\beta^\ell)^i \right) (\gamma^\ell)^j .
\end{equation}
We claim that $\sum_{i=0}^{n_1-1} (c_{i,j} (\beta^\ell)^i) =0$.

Since $\beta^{r_1}=1$ and $r_1$ divides $2^{n_1} -1$,
we have that $\beta \in \F_{2^{n_1}}$ and hence $\sum_{i=0}^{n_1-1} c_{i,j} (\beta^\ell)^i \in \F_{2^{n_1}}$, i.e., the
coefficient of $(\gamma^\ell)^j$ in Equation~(\ref{eq:for 2Dprac}) is an element of $\F_{2^{n_1}}$.

Let $m$ be the smallest integer such that $(\gamma^\ell)^m=\alpha^{\mu r_1 \ell m} =\delta^{\mu r_1 m} =1$.
Since the order of $\delta$ is $r_1 r_2$, it follows that $r_1 r_2$ divides $\mu r_1 m$. Moreover, $\mu r_1 + \nu r_2=1$ and
hence $\gcd (\mu,r_2)=1$ which implies that $r_2$ divides $m$, and therefore $r_2 \leq m$. However,
$(\gamma^\ell)^{r_2}=\alpha^{\mu r_1 \ell r_2} =\delta^{\mu r_1 r_2} =1$ and since $m$ is the smallest integer for which
$(\gamma^\ell)^m=1$, it follows that $m=r_2$ which implies that $\gamma^\ell$ has order $r_2$.
We claim that
$$
\gamma^\ell, (\gamma^\ell)^{2^{n_1}}, (\gamma^\ell)^{2^{2n_1}}, \ldots , (\gamma^\ell)^{2^{(n_2 -1)n_1}}
$$
are all distinct.

To see this, suppose for a contradiction that they are not distinct. Then
there exists $1 \leq e \leq n_2 -1$ such that $(\gamma^\ell)^{2^{n_1e}-1}=1$.
Since $\gamma^\ell=\alpha^{\mu r_1 \ell}=\delta^{\mu r_1}$,
$\gcd(r_2,\mu)=1$, and the order of $\delta$ is $r_1 r_2$, it follows that $r_1 r_2$ divides $\mu r_1 (2^{n_1 e}-1)$, so
$r_2$ divides $2^{n_1 e}-1$. Let $m$ be the smallest integer such that $\delta^{2^m-1}=1$. Since the order of $\delta$ is $r_1 r_2$
we have that $r_1 r_2$ divides $2^m -1$. Since $r_2$ divides $2^{n_1 e}-1$ and $r_1$ divides $2^{n_1}-1$, it follows that
$r_1$ divides $2^{n_1 e}-1$ and hence $2^{n_1 e}-1 =r_1 r_2 x$, for a positive integer $x$. Therefore, $m \leq n_1e$ and hence
$g(x)$ has fewer than $n_1 n_2$ distinct roots in $\F_{2^{n_1 n_2}}$, a contradiction. Hence the claim follows.
This implies that
$$
\gamma^\ell, (\gamma^\ell)^{2^{n_1}}, (\gamma^\ell)^{2^{2n_1}}, \ldots , (\gamma^\ell)^{2^{(n_2 -1)n_1}}
$$
are distinct, and $(\gamma^\ell)^{2^{n_1n_2}}= \gamma^\ell$.
Since $(\gamma^\ell)^{2^{n_1n_2}}=\gamma^\ell$, it follows that
$$
\prod_{i=0}^{n_2-1} (x - (\gamma^\ell)^{2^{i n_1}})
$$
is the minimal polynomial of $\gamma^\ell$ in $\F_{2^{n_1}}$. This polynomial has degree~$n_2$ in~$\F_{2^{n_1}}$.
However, the polynomial in Equation~(\ref{eq:for 2Dprac}) is a polynomial in $\gamma^\ell$ of smaller degree $n_2-1$ which
is equal to \emph{zero}.
This implies that all the coefficients of $(\gamma^\ell)^j$ in Equation~(\ref{eq:for 2Dprac}) are equal to \emph{zero}.
Therefore, for each $0 \leq j \leq n_2 -1$ we have
$$
\sum_{i=0}^{n_1-1} c_{ij} (\beta^\ell)^i =0 .
$$
In a similar way we can prove that $r_1$ is the smallest positive integer $m$ such that $(\beta^\ell)^m=1$, i.e., $\beta^\ell$ has order~$r_1$.
Since the $n_1$ integers $2^i ~ (\mmod ~ r_1)$ for $0 \leq i \leq n_1 -1$ are distinct, we have that
$$
\beta^\ell, (\beta^\ell)^{2}, (\beta^\ell)^{2^2}, \ldots , (\beta^\ell)^{2^{n_1-1}}
$$
are distinct, and $(\beta^\ell)^{2^{n_1n_2}}= \beta^\ell$. Therefore, the minimal polynomial of $\beta^\ell$ over $\F_2$ has
degree $n_1$ which implies that $c_{ij}=0$ for $0 \leq i \leq n_1 -1$ and $0 \leq j \leq n_2 -1$.

Therefore, the $n_1 \times n_2$ matrix in the upper-left corner of $\cB$ is a nonzero matrix and folding the sequences
generated by $g(x)$ yields an $(r_1,r_2;n_1,n_2)$-PRAC.
\end{IEEEproof}

\emph{Remark:}
The three proofs (Theorem~\ref{thm:PRA_MS} proved in~\cite{McSl76}, Theorem~\ref{thm:HuiminGen}, and Theorem~\ref{thm:HuiPRAC})
for the correctness of the constructions of PRAs using folding are very similar, but each one has different delicate
points in the proof associated with the conditions required for the construction.

\begin{example}
\label{ex:r2_notdivide_n2}
Let $n_1=3$, $n_2=4$, $r_1=7$, and $r_2=13$ and consider
the irreducible polynomial $g(x)= x^{12}+x^{10}+x^9+x+1$ with degree 12 and exponent 91.
This polynomial, $r_i$ and $n_i$, $i=1,2$, satisfy the conditions of Theorem~\ref{thm:HuiPRAC}
(because $7 \cdot 13=91$ divides $2^{3 \cdot 4}-1$, $\gcd(7,13)=1$, $7$ divides $2^3 -1$, and $2^0-1$, $2^1=2$, $2^2 =4$ are all distinct modulo $7$)
and hence folding the sequences generated by $g(x)$ into $7 \times 13$ arrays yields a $(7,13;3,4)$-PRAC.

There are 6 different irreducible polynomials of degree 12 and exponent 91, each one of them yields a $(7,13;3,4)$-PRAC.
\hfill\quad $\blacksquare$
\end{example}

\section{Analysis of the Constructions}
\label{sec:analysis}

We divide polynomials into three {\bf types}: primitive, INP and reducible.
In this section, the construction of PRACs from Section~\ref{sec:irreducible}, namely by folding the sequences generated by $g(x)=f_1(x)\vee f_2(x)$, is investigated. Which combinations of types of the polynomials $f_1(x)$, $f_2(x)$ and $g(x)$ are possible?
The results of the section are summarized in Table~\ref{tab:type} at the end of the
section: we provide examples of all combinations that do occur, and show why all other combinations cannot occur.
We also write down several families of PRACs which arise from the construction in~\ref{sec:irreducible}.

To understand which combinations of these types of three polynomials are possible we will have to use
Proposition~\ref{thm:unique_expo} and Corollary~\ref{cor:unique_expo}.
These results enable us to decide which combinations
of the three type of polynomials for $f_1(x)$, $f_2(x)$ and $g(x)$ are not possible.
As an example the first lemma has many different proofs and we have decided on one of them.

\begin{lemma}
\label{lem:g_no_pri}
The polynomial $g(x)=f_1(x) \vee f_2(x)$ in Theorem~\ref{thm:main_res} and Corollary~\ref{cor:main_res1} cannot be a primitive polynomial.
\end{lemma}
\begin{IEEEproof}
If $g(x)$ is a primitive polynomial, then it generates exactly one nonzero sequence.
Let $f_i(x)$, $i=1,2$, be a polynomial of degree $n_i$ and exponent $r_i$.
The number of nonzero sequences generated by $f_i(x)$ is $\frac{2^{n_i} -1}{r_i}$ which is at least 1.
The degree of $g(x)$ is $n_1 n_2$ and its exponent is $r_1 r_2$.
The number of nonzero sequences generated by $g(x)$ is
$$
\frac{2^{n_1 n_2}-1}{r_1 r_2} > \frac{2^{n_1} -1}{r_1} \frac{2^{n_2} -1}{r_2} \geq 1 ~.
$$
Therefore, $g(x)$ is not a primitive polynomial.
\end{IEEEproof}

While Lemma~\ref{lem:g_no_pri} asserts that $g(x)$ cannot be a primitive polynomial, the next lemma asserts that $g(x)$
is a reducible polynomial if $f_1(x)$ or $f_2(x)$ is a reducible polynomial
\begin{lemma}
\label{lem:suff_red}
If  $f_1(x)$ or $f_2(x)$ is a reducible polynomial, then the polynomial $g(x)=f_1(x) \vee f_2(x)$ in Theorem~\ref{thm:main_res},
is a reducible polynomial.
\end{lemma}
\begin{IEEEproof}
Let $f_i(x)$, $i=1,2$, be a polynomial of degree $n_i$ and exponent $r_i$. The polynomial $g(x)=f_1(x) \vee f_2(x)$
has degree $n_1 n_2$ and exponent $r_1 r_2$. The nonzero sequences generated by $g(x)$ have least period $r_1 r_2$ and they are
folded into $r_1 \times r_2$ arrays to form an $(r_1,r_2;n_1,n_2)$-PRAC with $\frac{2^{n_1 n_2}-1}{r_1 r_2}$ codewords.
Without loss of generality assume that $f_1(x)$ is a reducible polynomial. By Corollary~\ref{cor:unique_expo} there exists
an irreducible polynomial~$f_3(x)$ with exponent $r_1$ which is a proper factor of $f_1(x)$ and whose degree $n_3$ is smaller than $n_1$.
The polynomial $f_3(x) \vee f_2(x)$ also has exponent~$r_1 r_2$, it divides $g(x)$ but its degree $n_3 n_2$ is smaller than $n_1 n_2$, the degree of $g(x)$,
and hence again by using Corollary~\ref{cor:unique_expo} we have that $g(x)$ is a reducible polynomial.
\end{IEEEproof}

\begin{example}
\label{ex:irr_red_red}
Let $f_{1}(x) = \sum_{i=1}^{6} x^{i} + 1 = (x^3 + x + 1)(x^3 + x^2 + 1)$
be a reducible polynomial with exponent 7 and
$f_{2}(x)=\sum_{i=1}^{10} x^{i}+1$
be an INP polynomial with exponent~11.

$f_{1}(x) \vee f_{2}(x) = (x^{30} + x^{28} + x^{27} + x^{26} + x^{23} + x^{21} + x^{20} + x^{19} + x^{16} + x^{14} + x^{13} + x^{12} + x^{9} + x^8 + x^7 + x^4 + x^2 + x + 1)(x^{30} + x^{29} + x^{28} + x^{26} + x^{23} + x^{22} + x^{21} + x^{18} + x^{17} + x^{16} + x^{14} + x^{11} + x^{10} + x^9 + x^7 + x^4 + x^3 + x^2 + 1)$.
These two factors of $f_{1}(x) \vee f_{2}(x)$ are INP polynomials with exponent $77 = 7 \cdot 11$ and therefore
$f_{1}(x) \vee f_{2}(x)$ forms a ZF$(60,77)$.
By Theorem~\ref{thm:main_res} and Corollary~\ref{cor:main_res1}, folding its sequences into $7 \times 11$ arrays produces a $(7,11;6,10)$-PRAC.
\hfill\quad $\blacksquare$
\end{example}

\begin{example}
\label{ex:pri_red_red}
Let $f_1(x) = \sum_{i=0}^6 x^i = (x^3 + x^2 +1)(x^3 +x+1)$ be a reducible polynomial with exponent~7 and
let $f_2(x) = x^2 + x + 1$ be a primitive polynomial with exponent 3.
The polynomial
$$
f_1(x) \vee f_2(x) =(x^6+x^4+x^2+x+1)(x^6+x^5+x^4+x^2+1)
$$
is a reducible polynomial with exponent $21$.
By Theorem~\ref{thm:main_res} and Corollary~\ref{cor:main_res1}, folding its sequences into $7 \times 3$ arrays produces a $(7,3;6,2)$-PRAC.
\hfill\quad $\blacksquare$
\end{example}

For the next combinations of polynomial types for $f_1(x)$, $f_2(x)$ and $g(x)$, the following simple well-known lemma
is required.

\begin{lemma}
\label{lem:equivEvD}
Let $r_i$ and $n_i$, $i=1,2$, be positive integers such that $\gcd(r_1,r_2)=1$ and
\begin{equation}
\label{eq:equivEvD}
2^{n_i} \equiv 1 ~ (\mmod ~ r_i) .
\end{equation}
If $\ell = \gcd(n_1,n_2)$ then
\begin{equation*}
\label{eq:EXTequivEvD}
2^{n_1n_2/\ell} \equiv 1 ~ (\mmod ~ r_1 r_2).
\end{equation*}
If $n_1$ and $n_2$ are the smallest integers which satisfy Equation~(\ref{eq:equivEvD}), then $n_1n_2/\ell$ is the smallest value of~$k$ which
satisfies $2^k \equiv 1 ~ (\mmod ~ r_1 r_2)$.
\end{lemma}

%

\begin{lemma}
\label{lem:gcd_red}
If $f_i(x)$, $i=1,2$, are polynomials of degree $n_i$ and exponent $r_i$, such that $\gcd(r_1,r_2)=1$ and $\gcd(n_1,n_2)>1$, then
the polynomial $g(x)=f_1(x) \vee f_2(x)$ in Theorem~\ref{thm:main_res}, is a reducible polynomial.
\end{lemma}
\begin{IEEEproof}
Let $f_i(x)$, $i=1,2$, be polynomials of degree $n_i$ and exponent $r_i$. The polynomial $g(x)=f_1(x) \vee f_2(x)$
has degree $n_1 n_2$ and exponent $r_1 r_2$. The nonzero sequences generated by $g(x)$ have least period $r_1 r_2$ and they are
folded into $r_1 \times r_2$ arrays to form an $(r_1,r_2;n_1,n_2)$-PRAC with $\frac{2^{n_1 n_2}-1}{r_1 r_2}$ codewords.
By Lemma~\ref{lem:suff_red} if $f_1(x)$ or $f_2(x)$ is reducible, then $g(x)$ is a reducible polynomial and the lemma follows.
Hence, we may assume that both $f_1(x)$ and $f_2(x)$ are irreducible polynomials.
By Proposition~\ref{thm:unique_expo} $2^{n_1} \equiv 1 ~(\mmod ~ r_1)$ and $2^{n_2} \equiv 1 ~(\mmod ~ r_2)$.
By Proposition~\ref{thm:unique_expo} if $g(x)$ is an irreducible polynomial, then
\begin{equation}
\label{eq:on_irr_red}
2^{n_1 n_2} \equiv 1 ~ (\mmod ~ r_1 r_2)
\end{equation}
and $n_1n_2$ is the multiplicative order of 2 modulo $r_1 r_2$.
But, since $\ell=\gcd(n_1,n_2)>1$, it follows from Lemma~\ref{lem:equivEvD} that
$2^{n_1n_2/\ell} \equiv 1 ~ (\mmod ~ r_1 r_2)$ and hence $n_1n_2$ is not the multiplicative order of 2 modulo $r_1 r_2$.
Therefore, $g(x)$ is not an irreducible polynomial, i.e., it is a reducible polynomial.
\end{IEEEproof}

\begin{corollary}
\label{cor:gcd_irr_red}
If $f_i(x)$, $i=1,2$ are polynomials of degree $n_i$ with exponent $r_i$, such that $\gcd(n_1,n_2)>1$ and
$f_1(x)$ or $f_2(x)$ is an INP polynomial, then the polynomial $g(x)=f_1(x) \vee f_2(x)$
in Theorem~\ref{thm:main_res} and Corollary \ref{cor:main_res1} is a reducible polynomial.
\end{corollary}

\begin{example}
\label{ex:irr_irr_red}
Let $f_1(x) = x^4 +x^3 + x^2 + x + 1$ be an INP polynomial with exponent 5 and
$f_2(x) = x^6 + x^3 +1$ be an INP polynomial with exponent 9.
The polynomial
$$
f_1(x) \vee f_2(x) = x^{24} + x^{21} + x^{15} + x^{12} + x^9 + x^3  + 1=(x^{12}+x^9+1)(x^{12}+x^3+1)
$$
is a reducible polynomial with exponent $45$ which forms a ZF$(24,45)$.
By Theorem~\ref{thm:main_res} and Corollary \ref{cor:main_res1}, folding the sequences of this zero factor into $5 \times 9$ arrays produces
a $(5,9;4,6)$-PRAC.
\hfill\quad $\blacksquare$
\end{example}

\begin{example}
\label{ex:pri_irr_red}
Let $f_{1}(x) = x^4+x^3+x^2+x+1$
be an INP polynomial with exponent 5 and
$f_{2}(x)= x^6 + x^5 +1$
be a primitive polynomial with exponent 63.
The polynomial
$$
f_{1}(x) \vee f_{2}(x) = (x^{12} + x^4 + x^2 + x + 1) (x^{12} + x^{11} + x^{10} + x^9 + x^8 + x^6 + x^3 + x + 1).
$$
has two factors.
These two factors of $f_{1}(x) \vee f_{2}(x)$ are irreducible polynomials with exponent $315 = 5 \cdot 63$ and therefore
$f_{1}(x) \vee f_{2}(x)$ forms a ZF$(24,315)$.
By Theorem~\ref{thm:main_res} and Corollary \ref{cor:main_res1}, folding its sequences into $5 \times 63$ arrays produces a $(5,63;4,6)$-PRAC.
\hfill\quad $\blacksquare$
\end{example}

\begin{corollary}
\label{cor:Rprimes_irr}
If $f_i(x)$, $i=1,2$, are irreducible polynomial of degree $n_i$ and exponent $r_i$ such that $\gcd(n_1,n_2)=1$,
then the polynomial $g(x)=f_1(x) \vee f_2(x)$ in Theorem~\ref{thm:main_res} and Corollary \ref{cor:main_res1}, is an INP polynomial.
\end{corollary}
\begin{IEEEproof}
Since $f_i(x)$ is an irreducible polynomial, it follows from Proposition~\ref{thm:unique_expo} that $2^{n_i} \equiv 1 ~(\mmod ~ r_i)$ and
$n_i$ is the smallest integer which satisfies this equation. By Lemma~\ref{lem:equivEvD} we have that $k=n_1 n_2$ is the smallest integer
that satisfies the equation
\begin{equation}
\label{eq:small_kIRR_RED}
2^k \equiv 1 ~ (\mmod ~ r_1 r_2)~.
\end{equation}
Assume for a contradiction that $g(x)$ is a reducible polynomial of degree $n_1 n_2$ and exponent $r_1 r_2$.
It follows by Corollary~\ref{cor:unique_expo} that there exists an irreducible polynomial $h(x)$ which divides
$g(x)$ whose degree $n_3$ divides $n_1 n_2$ and its exponent is $r_1 r_2$, Hence there exists an integer smaller than $n_1 n_2$
which is a solution for $k$ in Equation~(\ref{eq:small_kIRR_RED}), which is a contradiction. Therefore, $g(x)$ is
an irreducible polynomial and by Lemma~\ref{lem:g_no_pri} it cannot be a primitive polynomial as required.
\end{IEEEproof}

\begin{example}
\label{ex:irr_irr_irr}
Let $f_{1}(x) = x^4+x^3+x^2+x+1$
be an INP polynomial with exponent 5 and
$f_{2}(x)= x^9 + x +1$ be an INP polynomial with exponent 73.
The polynomial
$$
f_{1}(x) \vee f_{2}(x) = x^{36} + x^{28} + x^{27} + x^{20} + x^{18} + x^{12} + x^{10} + x^9 + x^4 + x^3 + x^2 + x + 1
$$
is an INP polynomial with exponent $365 = 5 \cdot 73$.
By Theorem~\ref{thm:main_res} and Corollary \ref{cor:main_res1}, folding its sequences into $5 \times 73$ arrays yields a $(5,73;4,9)$-PRAC.
\hfill\quad $\blacksquare$
\end{example}


\begin{example}
\label{ex:irr_pri_irr}
Let $f_1(x) = \sum_{i=1}^4 x^i$ be an INP polynomial with exponent 5 and
Let $f_2(x) = x^3 + x^2 + 1$ be a primitive polynomial with exponent~7.
The polynomial
$$
f_{1}(x) \vee f_{2}(x) = x^{12} + x^{11} + x^{10} + x^{8} + x^{5} + x^{4} + x^{3} + x^{2} + 1
$$
is an INP polynomial with exponent $35 = 7 \cdot 5$.
By Theorem~\ref{thm:main_res} and Corollary \ref{cor:main_res1}, folding its sequences into $5 \times 7$ arrays yields a $(5,7;4,3)$-PRAC.
\hfill\quad $\blacksquare$
\end{example}

How many irreducible polynomials of degree $n$ and exponent $e$ exist? Some information on this question is given in the following results.
The first result can be found in~\cite[p. 41]{Gol67}.

\begin{lemma}
\label{lem:numIrrExp}
If $e$ is a factor of $2^n-1$ which is not a factor of any number $2^\ell -1$, where $\ell <n$, then there
are exactly $\frac{\phi(e)}{n}$ irreducible polynomials with exponent $e$, where $\phi$ is the Euler totient function.
\end{lemma}

\begin{lemma}
\label{lem:prodTwoPri}
If there are $k_i$ primitive polynomials of degree $n_i$, $i=1,2$, and ${\gcd (2^{n_1}-1,2^{n_2}-1)=1}$, then there are $k_1 k_2$
irreducible polynomials with exponent $(2^{n_1}-1)(2^{n_2}-1)$.
\end{lemma}
\begin{IEEEproof}
By Lemma~\ref{lem:numIrrExp} there are $k_1 = \frac{\phi(2^{n_1}-1)}{n_1}$ primitive polynomials of degree $n_1$ and exponent $2^{n_1}-1$.
By Lemma~\ref{lem:numIrrExp} there are $k_2 = \frac{\phi(2^{n_2}-1)}{n_2}$ primitive polynomials of degree $n_2$ and exponent $2^{n_2}-1$.
By Lemma~\ref{lem:numIrrExp} there are $\frac{\phi((2^{n_1}-1)(2^{n_2}-1))}{n_1n_2}$ irreducible polynomials of degree $n_1n_2$ and
exponent $(2^{n_1}-1)(2^{n_2}-1)$. Now,
$$
\frac{\phi((2^{n_1}-1)(2^{n_2}-1))}{n_1n_2}= \frac{\phi(2^{n_1}-1)}{n_1} \frac{\phi(2^{n_2}-1)}{n_2} = k_1k_2 .
$$
\end{IEEEproof}

Lemma~\ref{lem:prodTwoPri} implies each polynomial of degree $n_1 n_2$ and exponent $(2^{n_1}-1)(2^{n_2}-1)$ is
obtained from two distinct primitive polynomials of degrees $n_1$ and $n_2$ via $f_1(x) \vee f_2(x)$.

\begin{example}
\label{ex:two_primitive_irr}
Consider primitive polynomials $f_1(x)$ and $f_2(x)$ of degree 3 and 4, respectively, with exponents 7 and 15, respectively.

If $f_1(x) = x^4+x+1$ and $f_2(x)=x^3+x+1$ then
$$
f_1(x) \vee f_2(x) = x^{12} + x^9 + x^5 + x^4 + x^3 + x + 1.
$$

If $f_1(x) = x^4+x+1$ and $f_2(x)=x^3+x^2+1$ then
$$
f_1(x) \vee f_2(x) = x^{12} + x^8 + x^6 + x^5 + x^3 + x^2 + 1.
$$

If $f_1(x) = x^4+x^3+1$ and $f_2(x)=x^3+x+1$ then
$$
f_1(x) \vee f_2(x) = x^{12} + x^{10} + x^9 + x^7 + x^6 + x^4 + 1.
$$

If $f_1(x) = x^4+x^3+1$ and $f_2(x)=x^3+x^2+1$ then
$$
f_1(x) \vee f_2(x) = x^{12} + x^{11} + x^9 + x^8 + x^7 + x^3 + 1.
$$

In each of these four cases, $f_1(x) \vee f_2(x)$
is an INP polynomial with exponent $105 = 15 \cdot 7$ that forms a ZF$(12,105)$.
By Theorem~\ref{thm:main_res} and Corollary \ref{cor:main_res1}, folding the sequences of this zero factor into
$15 \times 7$ arrays is an $(15,7;4,3)$-PRAC.
\hfill\quad $\blacksquare$
\end{example}

Lemma~\ref{lem:prodTwoPri} can be generalized as follows.

\begin{lemma}
Assume there exist $k_i$ irreducible polynomials of degree $n_i$ with exponent $r_i$, $i=1,2$.
Assume further that $\gcd (r_1,r_2)=1$ and $\gcd (n_1,n_2)=\ell$. Then, there are exactly $\ell {k_1 k_2}$
irreducible polynomials of degree $\frac{n_1 n_2}{\ell}$ and exponent $r_1 r_2$. These polynomials can be partitioned
into $k_1 k_2$ sets of size $\ell$. Each such set is associated with exactly one pair of polynomials $(f_1(x),f_2(x))$,
such that $f_i$ has degree $n_i$, $i=1,n$ and $f_1(x) \vee f_2(x)$ is the product of the polynomials in the set.
\end{lemma}
\begin{IEEEproof}
Since $n_{i}$ is the smallest integer such that $2^{n_{i}} \equiv 1 \bmod r_{i} $ for $i=1,2$ and $\gcd(r_{1}, r_{2})=1$,
by Lemma~\ref{lem:equivEvD}, we know that $n=\frac{n_{1}n_{2}}{\ell}$ is the smallest integer such that $2^{n} \equiv 1 \bmod r_{1}r_{2}$.
Then by Lemma~\ref{lem:numIrrExp}, we know that there are $\frac{\phi(r_{1}r_{2})}{n}$ irreducible polynomials with exponent $r_{1}r_{2}$.
In addition, by Proposition~\ref{thm:unique_expo}, all these irreducible polynomials with exponents $r_{1}r_{2}$ have degree $n$.
Therefore, there are $\frac{\phi(r_{1}r_{2})}{n}$ irreducible polynomials with exponent $r_{1}r_{2}$ and degree $n$.
Now it remains to show that  $\frac{\phi(r_{1}r_{2})}{n} = \ell k_{1} k_{2}$.
Note that by Lemma~\ref{lem:numIrrExp}, $\phi(r_{i}) = n_{i} \cdot k_{i}$ for $i=1,2$.
We have $ \frac{\phi(r_{1}r_{2})}{n} = \frac{\phi(r_{1})\phi(r_{2})}{n}
= \frac{n_{1} k_{1} n_{2} k_{2}}{\frac{n_{1}n_{2}}{\ell}}  = \ell k_{1} k_{2}$
and the first claim follows.

Let $P$ be the set of all these irreducible polynomials with exponent $r_{1}r_{2}$ and degree $n$, and $R$ be their roots in $\F_{2^{n}}$.
Note that $R$ consists of all elements in $\F_{2^n}$ with order $r_{1}r_{2}$.
Let $P_{i}$ be the sets of the $k_{i}$ irreducible polynomials with degree $n_i$ and uniform exponent $r_{i}$ in $\F_{2^{n_{i}}}$,
and $R_{i}$ be their roots, where $i=1,2$.

To show the second claim, we first prove that there is a bijective mapping $f$ from $R$ to $R_{1} \times R_{2}$,
that is every $\gamma \in R$ has the unique pair $(\alpha, \beta) \in R_{1} \times R_{2}$ with $r = \alpha \cdot \beta$.
Since $\gcd(r_{1}, r_{2})=1$, there exist integers $x$ and $y$ such that $x r_{1} + y r_{2} = 1$.
Let $\alpha = \gamma^{yr_{2}}$ and $\beta = \gamma^{xr_{1}}$.
Then $\gamma = \alpha \cdot \beta$. In addition,
since the order of $\gamma$ is $r_{1}r_{2}$, $\gcd(y, r_{1})=1$, and $\gcd(x, r_{2})=1$,
we have $\alpha \in R_{1}$ and $\beta \in R_{2}$.
We can easily check that this mapping is injective. Then
$|f(R)| = |R| = \ell k_{1}k_{2} \cdot n  = k_{1} n_{1} \cdot k_{2} n_{2} = |R_{1} \times R_{2}|$,
shows that $f$ is a bijection.

Recall from the definition $f_{1}(x) \vee f_{2}(x) = \prod_{\alpha \in R(f_{1}), \\ \beta \in R(f_{2})} (x - \alpha \cdot \beta)$,
where $R(g)$ denotes the roots of a polynomial $g$.
Then each root $\gamma$ of any $f_{1}(x) \vee f_{2}(x)$ is associated
with a pair $(\alpha, \beta) \in R_{1} \times R_{2}$ by $\gamma = \alpha \cdot \beta$.
Such pairs derived from the roots of all $f_{1}(x) \vee f_{2}(x)$ are distinct since
the roots of a polynomial in $P_{1} (P_{2})$ are distinct and the roots of distinct irreducible polynomials are disjoint.
Actually, the set of such pairs is equal to $R_{1} \times R_{2}$.
From the previous paragraph, we know that the roots of all $f_{1}(x) \vee f_{2}(x)$ are in $R$,
which means that
\begin{enumerate}
\item[(1)] all the roots are distinct,

\item[(2)] each root of $f_{1}(x) \vee f_{2}(x)$ is also a root of a polynomial in $P$.
\end{enumerate}
From (2), we know that $f_{1}(x) \vee f_{2}(x)$ is a product of some polynomials in $P$
and (1) implies that these polynomials are distinct.
Since each factor has $\frac{n_{1}n_{2}}{\ell}$ distinct roots and $f_{1}(x) \vee f_{2}(x)$
has $n_{1}n_{2}$ distinct roots, it follows that there are $\ell$ polynomial factors.
This implies that each $f_{1}(x) \vee f_{2}(x)$ is a product of $\ell$ distinct polynomials in $P$.
By (1), the $k_{1}k_{2}$ sets of irreducible factors for each $f_{1}(x) \vee f_{2}(x)$ are disjoint.
The number of irreducible factors from all sets is $k_{1}k_{2} \cdot \ell = |P|$, which means that the sets form a partition of $P$
which concludes the proof.
\end{IEEEproof}

\begin{table*}[htbp]
\centering
\begin{tabular}{|c|c|c|c|}
\hline
$f_1(x)$    & $f_2(x)$ & $g(x)$ & references          \\ \hline
reducible   & reducible  & reducible &   Lemma~\ref{lem:suff_red}, ~Example~\ref{ex:red_red_red} \\ \hline
reducible   & INP  & reducible &   Lemma~\ref{lem:suff_red}, ~Example~\ref{ex:irr_red_red} \\ \hline
reducible   & primitive  & reducible &  Lemma~\ref{lem:suff_red}, ~Example~\ref{ex:pri_red_red}  \\ \hline
INP  & INP  & reducible & Corollary~\ref{cor:gcd_irr_red}, ~Example~\ref{ex:irr_irr_red}  \\ \hline
INP  & INP  & INP &  Corollary~\ref{cor:Rprimes_irr}, ~Example~\ref{ex:irr_irr_irr}  \\ \hline
INP   & primitive  & reducible & Corollary~\ref{cor:gcd_irr_red}, ~Example~\ref{ex:pri_irr_red}  \\ \hline
INP   & primitive  & INP &  Corollary~\ref{cor:Rprimes_irr}, ~Example~\ref{ex:irr_pri_irr}  \\ \hline
primitive   & primitive  & INP &  Corollary~\ref{cor:Rprimes_irr}, ~Example~\ref{ex:two_primitive_irr}  \\ \hline
\end{tabular}
\vspace{0.2cm}
\caption{Types of PRACs}
\label{tab:type}
\end{table*}

There are no more possible combinations of primitive, irreducible non-primitive, and reducible polynomials for $f_1(x)$, $f_2(x)$, and $g(x)$,
beside those which are demonstrated in the results and examples of this section.
These possible combinations are summarized in Table~\ref{tab:type}. The eight categories of the possible combinations partition
the PRACs which were constructed via Theorem~\ref{thm:main_res} and Corollary \ref{cor:main_res1} into equivalence classes.
They form a hierarchy between the various PRACs constructed so far.

We complete this part of the section with some specific infinite families of parameters for which there exist PRACs.
This list is only a drop in the sea from the possible parameters for such codes. These parameters were chosen since they look attractive to the authors.
However, many other sets of infinite families of parameters can be found relatively easy.

\begin{corollary}
\label{cor:two_primitive}
If $n_1$ and $n_2$ are two positive integers such that $\gcd(2^{n_1}-1,2^{n_2}-1)=1$, then there exists a $(2^{n_1}-1,2^{n_2}-1;n_1,n_2)$-PRAC.
\end{corollary}
\begin{IEEEproof}
The claim follows from the fact that there exist M-sequences of length $2^{n_1}-1$ and length $2^{n_2}-1$.
This implies that the conditions of Theorem~\ref{thm:main_res} and Corollary~\ref{cor:main_res1} are satisfied and the claim follows.
\end{IEEEproof}

\begin{corollary}
If $n_1$ and $n_2$ are two positive integers such that $\gcd(2^{n_1}-1,2^{n_2}-1)=1$ and there exist $k$ primitive polynomials
of degree $n_1$ and $\ell$ primitive polynomials of degree $n_2$, then there exists a $(2^{n_1}-1,2^{n_2}-1;kn_1,\ell n_2)$-PRAC.
\end{corollary}

\begin{corollary}
\label{cor:two_+1primes}
If $n_1+1$ and $n_2+1$ are distinct primes, then there exists an $(n_1+1,n_2+1;n_1,n_2)$-PRAC.
\end{corollary}
\begin{IEEEproof}
If $p$ is a prime, then the polynomial $\sum_{i=0}^{p-1} x^i$ has exponent $p$ and hence it forms
a ZF$(p-1,p)$. By definition $\gcd(n_1+1,n_2+1)=1$ which
implies that the conditions of Theorem~\ref{thm:main_res} and Corollary~\ref{cor:main_res1} are satisfied and the claim follows.
\end{IEEEproof}

\begin{corollary}
If $n_1$ and $n_2$ are two positive integers such that $\gcd(2^{n_1}-1,n_2+1)=1$ and $n_2+1$ is a prime,
then there exists a $(2^{n_1}-1,n_2+1;n_1,n_2)$-PRAC.
\end{corollary}

The following lemma yields a second hierarchy between the PRACs generated by multiplication of irreducible polynomials of the same degree
with the same exponent. It is an immediate consequence of the discussion in Lemma~\ref{lem:period}, Theorem \ref{thm:main_res}, and
Corollary \ref{cor:main_res1}.
\begin{lemma}
\label{lem:hier2}
Let $f_1(x)$ be an irreducible polynomial of degree $n_1$ and exponent $r_1$ and let $h_i (x)$, $1 \leq i \leq \ell$, $\ell \geq 2$ be
$\ell$ irreducible polynomials of degree $n_2$ and exponent $r_2$. The arrays of the $(r_1,r_2;n_1,\ell n_2)$-PRAC generated by folding
the sequences from $f_1(x) \vee \prod_{i=1}^{\ell} h_i(x)$ into $r_1 \times r_2$ arrays contain the
arrays of the $(r_1,r_2;n_1,(\ell-1)n_2)$-PRAC generated by folding
the sequences from $f_1(x) \vee \prod_{i=1}^{\ell-1} h_i(x)$ into $r_1 \times r_2$ arrays.
\end{lemma}

Unfortunately, Lemma~\ref{lem:hier2} does not give a complete picture of the hierarchy.
For example, there are eight irreducible polynomials of degree 8 and with exponent 85.
Folding the sequences generated by one of these polynomials yields a $(5,17;4,2)$-PRAC.
This PRAC is covered by Theorem~\ref{thm:HuiPRAC}.
Folding the sequences of the multiplication of two such polynomials yields a $(5,17;4,4)$-PRAC.
This PRAC is not covered by any theorem which was proved here.

\begin{conjecture}
\label{conj:contain}
Suppose that  $n_1 < r_1 < 2n_1$, and there are $\ell$ irreducible polynomials $f_1(x),f_2(x)\ldots f_\ell(x)$ of
degree $n_1n_2$ and exponent $r_1r_2$.
Suppose the set of sequences generated by each $f_i(x)$ forms an $(r_1,r_2; n_1, n_2)$-PRAC.
Folding the sequences generated by a product of
$k$ such polynomials, $1\leq k \leq\ell$, yields an $(r_1, r_2; n_1, k  n_2)$-PRAC.
\end{conjecture}

The condition $n_1 < r_1 < 2n_1$ in Conjecture~\ref{conj:contain} is necessary as demonstrated in the next example.

\begin{example}
Consider the two irreducible polynomials of degree 6 and exponent 21, $f_1(x)=x^6+x^5+x^4+x^2 +1$
and $f_2(x)=x^6+x^4+x^2+x+1$. The set of sequences generated by each one yields a $(3,7;2,3)$-PRAC.
Folding the product of these two polynomials yield a $(3,7;2,6)$-PRAC.

The set of sequences generated by each one also yields a $(7,3;3,2)$-PRAC, but now the condition $n_1 < r_1 < 2n_1$
is not satisfied for the product of these two polynomials.

There are six primitive polynomials of degree 6 and exponent 63. The sequence generated by each one yields a $(7,9;3,2)$-PRAC.
The condition $n_1 < r_1 < 2n_1$ is not satisfied for the product of these polynomials.
Consider the two primitive polynomials, $f_1(x)=x^6+x^5+1$ and $f_2(x)=x^6+x+1$. Folding the product of these two polynomials
does not yield a $(7,9;3,4)$-PRAC.
\hfill\quad $\blacksquare$
\end{example}

Finally, it is interesting to compare the PRACs constructed in Section~\ref{sec:irreducible},
which were extensively studied in the current section, and with the PRACs constructed in Section~\ref{sec:foldIrr}.

\begin{lemma}
\label{lem:compare_two}
The construction of PRACs using Theorem~\ref{thm:main_res} and Corollary~\ref{cor:main_res1}, where the folded sequences
were generated by an irreducible polynomial are generated also by Theorem~\ref{thm:HuiPRAC}.
\end{lemma}
\begin{IEEEproof}
The conditions in Theorem~\ref{thm:main_res} and Corollary~\ref{cor:main_res1} are that the sequences which are folded into
an $r_1 \times r_2$ array are generated by an irreducible polynomial $g(x)$ whose degree is $n_1 n_2$ and its exponent is $r_1 r_2$ and
the polynomial $f_1(x)$ has degree $n_1$ and exponent $r_1$.
By Table~\ref{tab:type} we have that $f_1(x)$ is an irreducible polynomial.
By Proposition~\ref{thm:unique_expo} we have that $n_1$ is the
smallest integer such that $2^{n_1} \equiv 1 ~ (\mmod ~ r_1)$. This implies that
$r_1$ divides $2^{n_1}-1$ and the integers $2^i ~(\mmod ~ r_1)$, $0 \leq i \leq n_1 -1$ are distinct
which are exactly the required conditions in Theorem~\ref{thm:HuiPRAC}.
\end{IEEEproof}

The PRACs produced using the construction implied by Theorem~\ref{thm:main_res} and Corollary~\ref{cor:main_res1}
in which $g(x)$ is a reducible polynomial cannot be constructed via
Theorem~\ref{thm:HuiPRAC}, since the sequences which are folded in Theorem~\ref{thm:HuiPRAC} are generated by an irreducible polynomial.
On the other hand, all other PRACs produced via Theorem~\ref{thm:main_res} and Corollary~\ref{cor:main_res1}
are also generated via Theorem~\ref{thm:HuiPRAC} as proved in Lemma~\ref{lem:compare_two}.
All the $(r_1,r_2;n_1,n_2)$-PRACs for which $r_1$ divides $2^{n_1}-1$ and $r_2$ does not divide $2^{n_2}-1$, which are generated
via Theorem~\ref{thm:HuiPRAC} (see Example~\ref{ex:r2_notdivide_n2}), are not generated via
Theorem~\ref{thm:main_res} and Corollary~\ref{cor:main_res1} since the definition of the polynomials
$f_1(x)$ and $f_2(x)$ requires these divisibility conditions when $g(x)$ is an irreducible polynomial.

\section{A Necessary and Sufficient Condition to Form a PRAC}
\label{sec:second_Necc_Suff}

In Section~\ref{sec:sufficient} we gave a necessary and sufficient condition that folding of sequences generated by an irreducible polynomial
form a PRA or a PRAC. In this section we provide a different necessary and sufficient condition which works also on sequences
generated from reducible polynomials.

Let $f_1(x),f_2(x),\ldots,f_k(x)$ be distinct irreducible polynomials over $\bF_2$ with degree $n$ and exponent~$e$.
Let $f(x)=\prod_{u=1}^k f_u(x)$. We are interested in finding a criterion to determine whether the folding of a set of sequences
with characteristic polynomial $f(x)$ produces a PRAC (or a PRA if the code contains exactly one codeword).
The criterion we provide can be checked in quadratic time, by finding the determinant of a certain binary matrix.
We represent the sequences with characteristic polynomial $f(x)$ in the following way:

\begin{lemma}
\label{lem:s_rep}
Let $f_1(x),f_2(x),\ldots,f_k(x)$ be distinct irreducible polynomials over $\bF_2$ of degree $n$. Let $f(x)=\prod_{u=1}^k f_u(x)$. For $1\leq u\leq k$, let $\alpha_u\in\bF_{2^n}$ be a root of $f_u(x)$. Let $t_1,t_2,\ldots,t_k:\bF_{2^n}\rightarrow \bF_2$ be non-trivial $\bF_2$-linear maps. Let $(s_i)$ be a sequence with characteristic polynomial $f(x)$. Then there are unique elements $\sigma_1,\sigma_2,\ldots ,\sigma_k\in\bF_{2^n}$ such that the terms of $(s_i)$ can be written in the form
\begin{equation}
\label{eqn:general_sum}
s_i=\sum_{u=1}^kt_u(\sigma_u\alpha_u^i)\text{ for }i\geq 0.
\end{equation}
\end{lemma}

We follow the approach in Rueppel~\cite[Section~3.3]{Rue86}), using the trace map. Recall that the \emph{trace map} from $\bF_{2^n}$ to $\bF_2$ is the function $\Tr:\bF_{2^n}\rightarrow \bF_2$ by $\Tr(x)=x+x^2+x^{2^2}+\cdots x^{2^{n-1}}$. The basic properties of this map are covered in Lidl and Niederreiter~\cite[Section~2.3]{LiNi97} (where the notation $\Tr_{\,\bF_{2^n}/\bF_2}$ is used for this map).
\begin{IEEEproof}
Let $(s_i)$ be a sequence with characteristic polynomial $f(x)$. Because $f_1(x),f_2(x),\ldots,f_k(x)$ are coprime polynomials, we may write
\[
(s_i)=\sum_{u=1}^k(s^u_i)
\]
for unique sequences $(s^u_i)$ where $(s_i^u)$ has characteristic polynomial $f_u(x)$. (This is a consequence of the properties of partial fractions, when we represent sequences as rational functions. See~\cite[Theorem~6.55]{LiNi97} and the discussion following it, for example.) By~\cite[Theorem~6.24]{LiNi97}, there exist unique elements $\theta_1,\theta_2,\ldots,\theta_k\in\bF_{2^n}$ such that $s^u_i=\Tr(\theta_u\alpha_u^i)$ for $i\geq 0$. Thus
\begin{equation}
\label{eqn:trace_sum}
s_i=\sum_{u=1}^k\Tr(\theta_u\alpha_u^i)
\end{equation}
for unique elements $\theta_1,\theta_2,\ldots,\theta_k\in\bF_{2^n}$.

By~\cite[Theorem~2.24]{LiNi97}, there are unique elements $\tau_1,\tau_2,\ldots,\tau_k\in\bF_{2^n}$ such that $t_u(z)=\Tr(\tau_u z)$ for $1\leq u\leq k$ and all $z\in\bF_{2^n}$. The elements $\tau_u$ are non-zero, as the maps $t_u$ are non-trivial. So~\eqref{eqn:trace_sum} implies that~\eqref{eqn:general_sum} holds
for unique elements $\sigma_1,\sigma_2,\ldots,\sigma_k\in\bF_{2^n}$, where $\sigma_u=\tau_u\theta_u$ for $1\leq u\leq k$.
\end{IEEEproof}

Let $r_1$ and $r_2$ be coprime positive integers. Let $\mu$ and $\nu$ be integers such that $\mu r_1+\nu r_2=1$. Let $f_1(x),f_2(x),\ldots f_k(x)$ be distinct irreducible polynomials of degree $n$ and exponent $e$, where $e=r_1r_2$. Let $n_1$ and $n_2$ be positive integers such that $kn=n_1n_2$. Let $\alpha_u \in\bF_{2^n}$ be a root of $f_u(x)$. Define $\beta_u=\alpha_u ^{\nu r_2}$ and $\gamma_u=\alpha_u^{\mu r_1}$. Let $t_1,t_2,\ldots,t_k:\bF_{2^n}\rightarrow \bF_2$ be non-trivial $\bF_2$-linear maps. For $1\leq u\leq k$, define the binary $n_1n_2\times n$ matrix $C_u$ as follows. The rows of $C_u$ are indexed by pairs $(i,j)$ where $0\leq i<n_1$ and $0\leq j<n_2$. The columns of $C_u$ are indexed by integers $v$ where $0\leq v<n$. Then the entry of $C_u$ in row $(i,j)$ and column $v$ is defined to be $t_u(\alpha_u^v\beta_{u}^i\gamma_u^j)\in\bF_2$. Finally, define the $n_1n_2\times n_1n_2$ matrix $C$ to be the concatenation $C=(C_1|C_2|\cdots|C_k)$ of the matrices $C_u$.

\begin{theorem}
\label{thm:det_check}
Let $r_1$ and $r_2$ be coprime positive integers. Let $f_1,f_2,\ldots f_k$ be distinct irreducible polynomials of degree $n$ and exponent $e$, where $e=r_1r_2$ and $n\geq 2$. Define $f(x)=\prod_{u=1}^kf_u(x)$. Let $n_1$ and $n_2$ be positive integers such that $n=n_1n_2/k$. Then the foldings of the non-zero sequences with characteristic polynomial dividing $f(x)$ form an $(r_1,r_2;n_1,n_2)$-PRAC if and only if $\det(C)\not=0$, where $C$ is the $n_1n_2\times n_1n_2$ matrix defined above.
\end{theorem}

\begin{IEEEproof}
We use the notation defined in the statement of the theorem, and also in the paragraph above the statement.

Suppose that $\det(C)=0$. Then there exists a non-zero (column) vector $\mathbf{d}$ such that $C\mathbf{d}=\mathbf{0}$. Write
\begin{equation}
\label{eqn:d_def}
\mathbf{d} =(d_{1,0},d_{1,1},\ldots,d_{1,n-1},d_{2,0},\ldots,d_{k,n-1})^T
\end{equation}
for some $d_{u,v}\in\bF_2$.  Define $\delta_u\in\bF_{2^n}$ by $\delta_u=\sum_{v=0}^{n-1}d_{u,v}\alpha_u^v$. Since $\mathbf{d}$ is a non-zero vector, not all the elements $\delta_u$ are zero. For all $i$ and $j$ with $0\leq i<n_1$ and $0\leq i<n_2$ we see that
\begin{equation}
\label{eqnentry}
\begin{split}\sum_{u=1}^kt_u(\delta_u \beta_u^i\gamma_u^j)&=\sum_{u=1}^k\sum_{v=0}^{n-1}t_u(d_{u,v} \alpha_u^v\beta_u^i\gamma_u^j)\\
&=\sum_{u=1}^k\sum_{v=0}^{n-1}d_{u,v}t_u(\alpha_u^v\beta_u^i\gamma_u^j)\\
&=0,
\end{split}
\end{equation}
the last equality following by the definition of entries of the row of $C$ indexed by $(i,j)$ and that fact that $C\mathbf{d}=\mathbf{0}$.

Choose non-zero elements $\sigma_1,\sigma_2,\ldots,\sigma_k\in\bF_{2^n}$ and $\sigma'_1,\sigma'_2,\ldots,\sigma'_k\in\bF_{2^n}$ such that $\sigma_u+\sigma'_u=\delta_u$ for $1\leq u\leq k$. Note that we can do this since $n\geq 2$. Moreover, note that $\sigma_u\not=\sigma'_u$ when $\delta_u$ is non-zero. Define binary sequences $(s_i)$ and $(s'_i)$ by
\begin{equation}
\label{eqn:sequence_form}
s_i=\sum_{u=1}^kt_u(\sigma_u \alpha_u^i)\text{ and }s'_i=\sum_{u=1}^kt_u(\sigma'_u \alpha_u^i).
\end{equation}
We see that the characteristic polynomials of $(s_i)$ and $(s'_i)$ each divide $f(x)$, since $f(\alpha_u)=0$ for all $u$. So the foldings of $(s_i)$ and $(s'_i)$ are shifts of elements in our array code. Note that $\sigma_u\not=\sigma'_u$ when $\delta_u$ is non-zero, so the sequences $(s_i)$ and $(s'_i)$ are not equal and their foldings are therefore not the same. The sum of the foldings of $(s_i)$ and $(s'_i)$ is the folding of $(s_i+s'_i)$. But the folding of $(s_i+s'_i)$ has an $n_1\times n_2$ all-zero window at its top-left corner, since for $0\leq i<n_1-1$ and $0\leq j<n_2 -1$ we see that the $(i,j)$ entry of the folding of $(s_i+s'_i)$ can be written as
\begin{align*}
\sum_{u=1}^kt_u(\sigma_u \beta_u^i\gamma_u^j)+\sum_{u=1}^kt_u(\sigma'_u \beta_u^i\gamma_u^j)&=\sum_{u=1}^kt_u((\sigma_u+\sigma'_u) \beta_u^i\gamma_u^j)\\
&=\sum_{u=1}^kt_u(\delta_u \beta_u^i\gamma_u^j)\\
&=0,\text{ by~\eqref{eqnentry}.}
\end{align*}
The foldings of $(s_i)$ and $(s'_i)$ cannot form part of an $(r_1,r_2;n_1,n_2)$-PRAC,
as they are distinct, but agree in their top left $n_1\times n_2$ window. So we do not have a PRAC, as required.

Now suppose that $\det(C)\not=0$. We will now show that the foldings of the non-zero sequences with characteristic polynomial dividing $f(x)$ form
an $(r_1,r_2;  n_1,n_2)$-PRAC. Let $(s_1)$ and $(s'_i)$ be two such sequences, and suppose that their foldings agree in an $n_1\times n_2$ window. Without loss of generality, by taking suitable shifts of our sequences, we may assume that the window is at the top left of the array. To show we have a PRAC, it suffices to show that $(s_i)=(s'_i)$.

Since $(s_i)$ and $(s'_i)$ have characteristic polynomial dividing $f(x)$, we may write them in the form~\eqref{eqn:sequence_form} for some elements $\sigma_u,\sigma'_u\in\bF_{2^n}$. Defining $\delta_u=\sigma_u+\sigma'_u$ for all $u$, we have
\[
s_i+s'_i=\sum_{u=1}^kt_u(\sigma_u \alpha_u^i)+\sum_{u=1}^kt_u(\sigma'_u \alpha_u^i)=\sum_{u=1}^kt_u(\delta_u \alpha_u^i).
\]
We may write $\delta_u=\sum_{v=0}^{n-1}d_{u,v}\alpha_u^v$ for some elements $d_{u,v}\in\bF_2$.
The top left $n_1\times n_2$ window of the folding of the sequence $(s_i+s'_i)$ is all zero, and hence for all $0\leq i<n_1$ and $0\leq j<n_2$
\begin{align*}
0&=\sum_{u=1}^kt_u(\delta_u \beta_u^i\gamma_u^j)\\
&=\sum_{u=1}^k\sum_{v=0}^{n-1}t_u(d_{u,v} \alpha_u^v\beta_u^i\gamma_u^j)\\
&=\sum_{u=1}^k\sum_{v=0}^{n-1}d_{u,v}t_u(\alpha_u^v\beta_u^i\gamma_u^j).
\end{align*}
Defining $\mathbf{d}$ using~\eqref{eqn:d_def}, we see that the $(i,j)$ entry of $C\mathbf{d}$ is $0$ for all $0\leq i<n_1$ and $0\leq j<n_2$, and so $\mathbf{d}$ lies in the null space of $C$. As $\det(C)\not=0$, the matrix $C$ has trivial null space and so $\mathbf{d}=0$. Thus $d_{u,v}=0$ for all $u$ and $v$, which implies that $\delta_{u}=0$ for $1\leq u\leq k$. Since $\sigma_u+\sigma'_u=\delta_u=0$, we see that $\sigma_u=\sigma'_u$ for $1\leq u\leq k$, and so $(s_i)=(s'_i)$ as required.
\end{IEEEproof}

Note that for any fixed $u$ with $1\leq u\leq k$, the elements $\alpha_u^v$ for $0\leq v<n$ are a basis for $\bF_{2^n}$ over $\bF_2$.
These elements occur in the definition of the matrix $C$, but they can be replaced by any other basis
for $\bF_{2^n}$ over $\bF_2$. The proof of the theorem is essentially the same if this change is made.

\vspace{0.5cm}

The following corollary (giving a `dual' condition to that in Theorem~\ref{thm:det_check} in the special case when $k=1$) is very close in spirit to Theorem~\ref{thm:HuiPRAC}.

\begin{corollary}
\label{cor:trace_check}
Let $r_1$ and $r_2$ be coprime positive integers. Let $\mu$ and $\nu$ be integers such that $\mu r_1+\nu r_2=1$. Let $f(x)$ be an irreducible polynomial of degree $n$ and exponent $e$, where $e=r_1r_2$ and $n\geq 2$. Let $n_1$ and $n_2$ be positive integers so that $n=n_1n_2$. Let $\alpha\in\bF_{2^n}$ be a root of $f(x)$. Define $\beta=\alpha^{\nu r_2}$ and $\gamma=\alpha^{\mu r_1}$. Regard the elements of $\bF_{2^n}$ as binary vectors of length $n$. Then the foldings of the non-zero sequences with characteristic polynomial $f(x)$ form an $(r_1,r_2;n_1,n_2)$-PRAC if and only if the following $n_1n_2=n$ vectors in $\bF_{2}^n$ are linearly independent:
\[
\beta^i\gamma^j\text{ where }0\leq i< n_1\text{ and }1\leq j<n_2.
\]
\end{corollary}
\begin{IEEEproof}
Let $C$ be the $n\times n$ matrix whose rows are indexed by pairs $(i,j)$ with $0\leq i<n_1$ and $0\leq j<n_2$, columns indexed by integers $v$ with $0\leq v<n$, where the entry in row $(i,j)$ and column $v$ is equal to
\[
c_{(i,j),v}=\Tr(\alpha^v\beta^i\gamma^j).
\]
By Theorem~\ref{thm:det_check} (in the case when $t_u=\Tr$ for $1\leq u\leq k$), our foldings fail to form a PRAC if and only if $\det(C)=0$.

Let $\mathbf{d}=(d_0,d_1,\ldots ,d_{n-1})^T$, with $d_v\in\bF_2$ for $0\leq v<n$, be a column vector. Let $\delta=\sum_{v=0}^{n-1}d_v\alpha^v$ be the corresponding element of $\bF_{2^n}$. For any pair $(i,j)$ with $0\leq i<n_1$ and $0\leq j<n_2$, we have that
\begin{align*}
\sum_{v=0}^{n-1}c_{(i,j),v}d_v&=\sum_{v=0}^{n-1}d_v\Tr(\alpha^v\beta^i\gamma^j)\\
&=\Tr\left(\left(\sum_{v=0}^{n-1}d_v\alpha^v\right)\beta^i\gamma^j\right)\\
&=\Tr(\delta\beta^i\gamma^j).
\end{align*}
Thus $\mathbf{d}$ lies in the (column) null space of $C$ if and only if
\begin{equation}
\label{eqn:trace}
\Tr(\delta\beta^i\gamma^j)=0\text{ for }0\leq i<n_1\text{ and }0\leq j<n_2.
\end{equation}

Define the $\bF_2$-linear map $t:\bF_{2^n}\rightarrow\bF_2$ by $t(z)=\Tr(\delta z)$ for all $z\in\bF_{2^n}$. Suppose the elements $\beta^i\gamma^j$ are linearly independent, so they form a basis for $\bF_{2^n}$. The conditions~\eqref{eqn:trace} then imply that $t$ is the trivial map. Hence, by~\cite[Theorem~2.24]{LiNi97}, we have $\delta=0$. So the null space of $C$ is trivial and therefore $\det(C)\not=0$. Hence our foldings form a PRAC.

If the elements $\beta^i\gamma^j$ are linearly dependent, the $n$ linear constraints~\eqref{eqn:trace} are dependent, and so there exist non-zero solutions $\delta$ to these constraints. Thus the null space of $C$ is non-trivial, so $\det(C)=0$, and hence our foldings do not form a PRAC.
\end{IEEEproof}

\section{Conclusion and Future Research}
\label{sec:conclude}

Pseudo-random arrays and pseudo-random array codes are constructed by folding sequences generated by irreducible polynomials,
or by reducible polynomials which are a product of distinct irreducible polynomials of the same degree and the same exponent.
In the current work these structures constructed by folding are examined.
Necessary and sufficient conditions to verify whether the array or the array code is a PRA or a PRAC, respectively, are given.
Several techniques that require simple computations for divisibility of integers to
verify if some of these structures are PRAs or PRACs are proved. One set of conditions which generalizes
the conditions given by MacWilliams and Sloane~\cite{McSl76} for generating one pseudo-random array
is shown to yield pseudo-random arrays not known before. This set of conditions was generalized for pseudo-random array codes.
Another completely different set of conditions with apparently a different construction for pseudo-random array codes
which can be also generated by folding is also presented. Finally, the codes constructed by the different constructions
are compared and two hierarchies implied by the constructions are discussed.

There are many other interesting problems which arise from our discussion. Some of them are as follows:

\begin{enumerate}
\item We did not consider constructions of $(r_1,r_2;n_1,n_2)$-SDBACs which are not PRACs and might have parameters not obtained
by PRACs. It is not difficult to construct them for $n_2=2$.
Beyond $n_2=2$ it is an intriguing problem to construct such array codes which are not PRACs.
This is a promising direction for future research.

\item Some of the constructed pseudo-random array codes have large minimum distance. This is mentioned with an example in~\cite{Etz24a}.
An analysis of the minimum distance of these codes is an interesting question. These codes are linear codes and it is interesting to
compare their parameters with other linear array codes.

\item Experimental results show that there exist more parameters for which folded sequences generated by an irreducible polynomial
produces a PRA or a PRAC beyond Theorem~\ref{thm:HuiminGen} and Theorem~\ref{thm:HuiPRAC}. Examples of such pseudo-random array codes
were presented in Section~\ref{sec:sufficient}.
We would like to see a comprehensive solution for this case, i.e., for which parameters and for
which M-sequences, folding the M-Sequences yields pseudo-random arrays; the same question for which parameters and for
which sequences generated by only-irreducible polynomial, folding yields pseudo-random array codes.

\item Experiments show that folding sequences generated by reducible polynomials whose factors are distinct irreducible polynomials
with the same degree and the same exponent generates a pseudo-random array.
Only a few cases are covered by Theorem~\ref{thm:main_res} and Corollary~\ref{cor:main_res1}.
We would like to see a comprehensive treatment of folding for such sequences.

\item We would like to see a proof of, or a counterexample for Conjecture~\ref{conj:contain} in Section~\ref{sec:analysis}.
\end{enumerate}

\section*{Acknowledgement}

The authors thank Ronny Roth for pointing on Theorem 8.67 in~\cite{LiNi97}.



\end{document}